\documentclass[fleqn,usenatbib,twocolumns]{mnras}

\usepackage{newtxtext,newtxmath}

\usepackage[T1]{fontenc}

\DeclareRobustCommand{\VAN}[3]{#2}
\let\VANthebibliography\thebibliography
\def\thebibliography{\DeclareRobustCommand{\VAN}[3]{##3}\VANthebibliography}

\defcitealias{bright2023}{Bright \& Rhodes et al. (2023)}

\usepackage{graphicx}	
\usepackage{amsmath}	
\usepackage{longtable}
\usepackage{subcaption}
\usepackage{graphicx}
\usepackage[normalem]{ulem}

\title[Radio observations of GRB 221009A]{Rocking the BOAT: the ups and downs of the long-term radio light curve for GRB 221009A }

\author[L. Rhodes et al.]
{L. Rhodes,$^{1}$\thanks{E-mail: lauren.rhodes@physics.ox.ac.uk}
A. J. van der Horst,$^{2}$,
J. S. Bright$^{1}$,
J. K. Leung,$^{3,4,5}$,
G. E. Anderson$^{6}$,
R. Fender$^{1,7}$,
\newauthor
J. F. Ag\"{u}\'{i} Fernandez$^{8}$, 
M. Bremer,$^{9}$ 
P. Chandra$^{10}$,
D. Dobie$^{11,12}$,
W. Farah$^{13,14}$,
S. Giarratana,$^{15}$,
K. Gourdji$^{16}$,
\newauthor
D. A. Green$^{17}$,
E. Lenc$^{18}$,
M.~J.~Micha{\l}owski$^{19}$,
T. Murphy$^{11,12}$, 
A. J. Nayana$^{20}$, 
A. W. Pollak$^{13,14}$,
\newauthor
A. Rowlinson$^{21,22}$,
F. Schussler$^{23}$,
A. Siemion$^{1,13,14,24,25}$
R. L. C. Starling$^{26}$,
P. Scott$^{17}$,
C. C. Th\"{o}ne$^{27}$, 
\newauthor
D. Titterington$^{17}$
A. de Ugarte Postigo$^{9,28}$\\
$^{1}$Astrophysics, Department of Physics, University of Oxford, Denys Wilkinson Building, Keble Road, Oxford, OX1 3RH, UK\\
$^{2}$Department of Physics, George Washington University, 725 21st St NW, Washington, DC, 20052, USA\\
$^{3}$David A. Dunlap Department of Astronomy and Astrophysics, University of Toronto, 50 St. George Street, Toronto, ON M5S 3H4, Canada\\
$^{4}$Dunlap Institute for Astronomy and Astrophysics, University of Toronto, 50 St. George Street, Toronto, ON M5S 3H4, Canada\\
$^{5}$Racah Institute of Physics, The Hebrew University of Jerusalem, Jerusalem 91904, Israel\\
$^{6}$International Centre for Radio Astronomy Research, Curtin University, GPO Box U1987, Perth, WA 6845, Australia\\
$^{7}$Department of Astronomy, University of Cape Town, Private
Bag X3, Rondebosch 7701, South Africa\\
$^{8}$Centro Astron\'{o}mico Hispano en Andaluc\'{i}a, Observatorio de Calar
Alto, Sierra de los Filabres, G\'{e}rgal, Almer\'{i}a, 04550, Spain\\
$^{9}$Observatoire de la C\^{o}te d’Azur, Universit\'{e} C\^{o}te d’Azur,
Boulevard de l’Observatoire, 06304 Nice, France\\
$^{9}$Institut de Radioastronomie Millimétrique, 300 Rue de la Piscine, 38400 Saint-Martin-d'Hères, France\\
$^{10}$National Radio Astronomy Observatory, 520 Edgemont Rd, Charlottesville VA 22903, USA\\
$^{11}$Sydney Institute for Astronomy, School of Physics, The University of Sydney, New South Wales 2006, Australia\\
$^{12}$ARC Centre of Excellence for Gravitational Wave Discovery (OzGrav), Hawthorn, Victoria, Australia\\
$^{13}$SETI Institute, 339 Bernardo Ave, Suite 200 Mountain View,
CA 94043, USA\\
$^{14}$Berkeley SETI Research Centre, University of California, Berkeley, CA 94720, USA\\
$^{15}$Istituto Nazionale di Astrofisica, Osservatorio Astronomico di Brera, via E. Bianchi 46, 23807 Merate, (LC), Italy\\
$^{16}$Centre for Astrophysics and Supercomputing, Swinburne University of Technology, P.O. Box 218, Hawthorn, VIC 3122, Australia\\
$^{17}$Astrophysics Group, Cavendish Laboratory, 19 J.J. Thomson Avenue, Cambridge, CB3 0HE, UK\\
$^{18}$Australian Telescope National Facility, CSIRO Astronomy and Space Science, PO Box 76, Epping, NSW 1710, Australia\\
$^{19}$Astronomical Observatory Institute, Faculty of Physics, Adam Mickiewicz University, ul. S\l{}oneczna 36, 60-286 Pozna\'{n}, Poland\\
$^{20}$Department of Astronomy, University of California, Berkeley, USA\\
$^{21}$Anton Pannekoek Institute for Astronomy, University of Amsterdam, Science Park 904, 1098 XH, Amsterdam, the Netherlands\\
$^{22}$ASTRON, the Netherlands Institute for Radio Astronomy, Oude Hoogeveensedĳk 4, 7991 PD Dwingeloo, the Netherlands\\
$^{23}$IRFU, CEA, Universit\'{e} Paris-Saclay, F-91191 Gif-sur-Yvette, France\\
$^{24}$Department of Physics and Astronomy, University of Manchester, UK\\
$^{25}$University of Malta, Institute of Space Sciences and Astronomy, Msida, MSD2080, Malta\\
$^{26}$University of Leicester, School of Physics and Astronomy, University Road, Leicester LE1 7RH\\
$^{27}$Astronomical Institute of the Czech Academy of Sciences
(ASU-CAS), Fricova 298, Ond\v{r}ejov, 251 65, Czech Republic\\
$^{28}$Aix Marseille Univ, CNRS, CNES, LAM, Marseille, France
}

\date{Accepted XXX. Received YYY; in original form ZZZ}

\pubyear{2015}

\begin{document}
\label{firstpage}
\pagerange{\pageref{firstpage}--\pageref{lastpage}}
\maketitle

\begin{abstract}
We present radio observations of the long-duration gamma-ray burst (GRB) 221009A which has become known to the community as the Brightest Of All Time or the BOAT. Our observations span the first 475\,days post-burst and three orders of magnitude in observing frequency, from 0.15 to 230\,GHz. By combining our new observations with those available in the literature, we have the most detailed radio data set in terms of cadence and spectral coverage of any GRB to date, which we use to explore the spectral and temporal evolution of the afterglow. By testing a series of phenomenological models, we find that three separate synchrotron components best explain the afterglow. 
The high temporal and spectral resolution allows us to conclude that standard analytical afterglow models are unable to explain the observed evolution of GRB 221009A. We explore where the discrepancies between the observations and the models are most significant and place our findings in the context of the most well-studied GRB radio afterglows to date. Our observations are best explained by three synchrotron emitting regions which we interpret as a forward shock, a reverse shock and an additional shock potentially from a cocoon or wider outflow. Finally, we find that our observations do not show any evidence of any late-time spectral or temporal changes that could result from a jet break but note that any lateral structure could significantly affect a jet break signature.
\end{abstract}




\section{Introduction}

Long-duration gamma-ray bursts (GRBs) are produced in highly relativistic jets, launched during the collapse of massive stars, and they are the most powerful explosions in the Universe. GRB 221009A has been dubbed the Brightest of all Time, or the BOAT \citep{burns2023}. Lasting about 600\,seconds, the variable, high energy, prompt emission was detected by the Neil Gehrels \textit{Swift} Observatory -- Burst Alert Telescope and X-ray telescope \citep[BAT and XRT, respectively, ][]{williams2023}, Insight-HXMT and GECAM-C \citep{2023arXiv230301203A}, Konus Wind and SRG/ART-X \citep{2023ApJ...949L...7F} and the Fermi Gamma-ray Space Telescope \citep{lesage2023}. Placed at a redshift of 0.151 \citep{2022GCN.32648....1D, 2023arXiv230207891M}, the isotropic gamma-ray energy output has been measured as $1\times10^{55}$\,erg, $1.5\times10^{55}$\,erg and $1.2\times10^{55}$\,erg by \citet[][between 1-10,000\,keV, 10\,keV -- 6\,MeV and 20\,keV -- 10\,MeV, respectively]{lesage2023, 2023arXiv230301203A, 2023ApJ...949L...7F}, nearly twice the value of the next most energetic, GRB 080916C \citep{2009A&A...498...89G}. Given its prompt emission properties, it has been established as a once in 10,000\,year event \citep{burns2023}. In fact, GRB 221009A was so bright that the prompt emission caused disturbances in the ionosphere \citep{2022RNAAS...6..222H}.

The afterglow to GRB 221009A has been detected consistently between 0.4\,GHz and 20\,TeV \citep{laskar2023, lhaaso2023}. In terms of spectral coverage, it exceeds the all other TeV afterglows with radio detections like GRB 190114C or GRB 190829A \citep{2019Natur.575..455M, 2021Sci...372.1081H}. In terms of data quantity and quality, it exceeds the GHz-to-GeV afterglow of GRB 130427A \citep[e.g. ][]{vanderhorst2014,2014ApJ...792..115L,2014Sci...343...42A,2016MNRAS.462.1111D}, although the latter had a much better sampling of optical light curves since it did not suffer from extinction in the way that GRB 221009A did \citep{Fulton2023, levan2023}. Similar to GRB 130427A \citep{2014MNRAS.440.2059A}, \citetalias{bright2023} showed that GRB 221009A had a bright light curve peak at 15~GHz within the first day, followed by an overall decline at radio frequencies. This behaviour is quite different from the `classical' well-sampled radio afterglows of, for instance, GRB 970508 and GRB 030329, which have peaks at timescales of weeks to months \citep{2000ApJ...537..191F,2005ApJ...634.1166V,2005A&A...440..477R}. The origin of the early time radio peaks are thought to be from reverse shock, produced from a shock front propagating back through the jet. Details of the light curve behaviour, in particular over a wide frequency range, give important insights into the underlying physics at various scales, from the jetted explosion outflow to the accelerated particles generating the observed emission \citep[e.g.,][]{sari1998,wijers1999}.

The focus of this paper is the radio emission from GRB 221009A, covering three orders of magnitude in both observing frequency and days post-burst. While the TeV emission leads to various questions regarding possible emission processes at these high energies and the potential for detecting GRBs at TeV energies more frequently \citep{2019Natur.575..455M,2021Sci...372.1081H,2024MNRAS.527.5856A}, the radio observations provide the necessary context for understanding the physics of the jetted GRB outflow, together with multi-wavelengths observations in the optical and X rays \citep[e.g.][]{2023MNRAS.524L..78G, oconnor2023}. The light curve behaviour of GRB 221009A in the first days to weeks do not seem to follow expectations of the standard model that is typically used to describe radio afterglows \citep[e.g.][]{wijers1999,granot2002}. The extremely dense sampling of the light curves at various radio frequencies as presented in this paper is unprecedented and allows for detailed modelling that will lead to better descriptions of GRB jets and the relevant emission processes.

The dominant emission mechanism in GRB afterglows at radio frequencies is synchrotron radiation from extremely relativistic electrons accelerated by shocks at the front of a relativistic collimated outflow \citep{1993ApJ...405..278M,sari1998}. This is also the emission mechanism assumed to be at play in the GRB 221009A afterglow. While we are only considering one emission mechanism, i.e., synchrotron, there can be multiple emission sites. For instance, the jet sweeping up particles in the ambient medium leads to a forward shock, but will also lead to the formation of the aforementioned reverse shock which can be dominant at early times given the right conditions \citep{2000ApJ...542..819K}. Besides this shock structure in the radial direction, there can also be structure in the lateral direction. This structure could be smooth, for instance, a structured energy profile as a function of distance to the jet axis instead of a homogeneous energy profile \citep{rossi2002,2021MNRAS.506.4163L, 2022Galax..10...93S}; but there could also be multiple jet components \citep{2005MNRAS.360..305S}, and potentially a cocoon around the jet \citep{ramirezruiz2002,2017ApJ...834...28N,2019Natur.565..324I}. This could lead to multiple synchrotron emission components, or emission components that evolve differently from the canonical top-hat behaviour \citep{vanderhorst2014,2019MNRAS.486.2721B, 2022MNRAS.513.1895R}.

Besides these macrophysical considerations, high-quality multi-wavelength data as presented here reveals nuances in the microphysics of GRB afterglows. Afterglow modelling can lead to insights into the magnetic field strength and energetics, but also the total energetics, acceleration efficiency, and energy distribution of the accelerated electrons \citep{granot2002,eichlergranot2005}. To complicate this further, detailed simulations of particle acceleration and magnetic field amplification by relativistic shocks indicate that there is potentially a time dependence of the energies in magnetic fields and electrons \citep{2003MNRAS.339..881R}, and this has also been adopted in multi-wavelength modelling of some GRB afterglows with peculiar behaviour \citep{vanderhorst2014,2019MNRAS.486.2721B, 2021MNRAS.504.5685M, 2022ApJ...931L..19S}. 

Given the extremely high quality of the radio data presented in this paper, and the dynamics of the synchrotron spectrum that is likely quite different from the standard behaviour, we take a fairly cautious approach in the modelling presented here. While a standard GRB synchrotron spectrum is still assumed, the temporal evolution of the spectrum is kept free of constraints where possible, to provide input on detailed modelling and theoretical efforts, and get a better handle on the interpretation of the wealth of these data from this unique source. We highlight here the use of the convention $F_{\nu} \propto t^{\alpha}\nu^{\beta} $ throughout this work to describe the temporal and spectral evolution. This paper is laid out in the following manner: in Section \ref{sec:obs} we present the new radio observations and the data reduction methods used; in Section \ref{sec:model}, we lay out the results of our observing campaigns and describe the model used to explain the data; in Section \ref{sec:discussion} we put our results in a broader context and interpret the data using various models; and we conclude in Section \ref{sec:conc}.

\section{Observations}\label{sec:obs}

Here we present the data reduction processes for the observations used in this work. The flux density measurements and upper limits for our new observations are summarised in Table \ref{tab:radio_obs}. In addition to the datasets we present here, our work also incorporates the previously published radio data from \citet{laskar2023, Giarratana2023} and \citetalias{bright2023}, and the X-ray data from \citet{williams2023}.

\subsection{AMI--LA}
The Arcminute Microkelvin Imager -- Large Array (AMI--LA) is an eight-dish interferometer based in Cambridge, UK \citep{2008MNRAS.391.1545Z}. It observes at a central frequency of 15.5\,GHz with a bandwidth of 5\,GHz, achieving an angular resolution of about 30\,arcsec \citep{2018MNRAS.475.5677H}. \citetalias{bright2023} presented the first five days of observations from AMI--LA, and here we present the rest of the observing campaign. We continued to observe the position of GRB 221009A almost daily until 210\,days post-burst when the first non-detection occurred. Between 210 and 320\,days post-burst, we concatenated separate non-detections to obtain deeper limits.

AMI--LA data is reduced using a custom software package: \textsc{reduce\_dc} \citep{2013MNRAS.429.3330P}. The software performs bandpass and flux scaling using 3C286 and complex gain calibration using J1925+2106. Flagging and imaging is done in \textsc{casa} using the tasks \textit{rflag}, \textit{tfcrop} and \textit{clean} \citep{McMullin2007}. The details of observing times and measured flux densities are provided in Table \ref{tab:radio_obs}. We note that unlike in \citetalias{bright2023}, we do not split each observation up, because the duration of a given epoch is a negligible fraction of the total time since the burst was first detected, so no significant evolution is expected within an observation.

\subsection{ASKAP}
We obtained target-of-opportunity observations of the GRB 221009A field with the Australian Square Kilometre Array Pathfinder \citep[ASKAP, ][]{Johnston2007}.
Our observations were centred on 888 MHz, with a bandwidth of 288 MHz, taken using the \texttt{square\_6x6} beam footprint \citep[see figure 20 of ][]{hotan2021}.
The data products for these observations can be found under the project code AS113 
with SBIDs: 44780, 44857, 44918, 45060, 45086, 45416, 46350, 46419, 46492, 46554 and 48611 in the CSIRO ASKAP Science Data Archive (CASDA\footnote{\url{https://research.csiro.au/casda/}}).

Observations of PKS B1934$-$638 were used to calibrate the antenna gains, bandpass and the absolute flux-density scale.
Flagging of radio frequency interference, calibration of raw visibilities, full-polarisation imaging, and source finding on total intensity images were all performed through the standard ASKAPsoft pipeline \citep{guzman2019}.
The resulting image reached a typical rms of ${\sim} 50\,\mu$Jy$\,$beam$^{-1}$.
We evaluated and corrected for the systematic flux-scale offset by comparing the flux density of field sources in each observation against those in the Rapid ASKAP Continuum Survey (RACS) catalogue \citep{hale2021}.

\subsection{ATA}
Located $\sim$\,200 miles north of San Fransisco, the Allen Telescope Array is a 42-element radio interferometer hosted at the Hat Creek Radio Observatory. 
Mounted on the focus of each element is a dual-polarization, log-periodic feed that is cryogenically cooled and sensitive to radiation in the range of 1 to 12 GHz. 
Analogue signals from the array are transmitted through fibre to a centralised signal processing room and are split into 4 independent chains that get multiplexed by 4 tunable local oscillators in a super-heterodyne system. 
The current correlator backend supports the digitisation of 2 out of the 4 available tunings for 20 of the 42 antennas, where each tuning can be placed anywhere in the available RF range of the log periodic feed, with $\sim 700$\,MHz of usable bandwidth for each.

The radio counterpart of GRB 221009A was observed extensively with the ATA  beginning just a few hours after the burst as reported in \citetalias{bright2023}. Here we build on that work and utilised the flexible frequency tunability of the ATA to monitor the $1$--$10\,\rm{GHz}$ spectral evolution over its entire outburst.  Either 3C147, 3C48, or 3C286 was observed as flux calibrator at the beginning of each observing block, and a 10\,minute observation of the phase calibrator J1925$+$2106 was interleaved for every 30\,minutes of science target recording (regardless of observing frequency). We evolved our total integration time on source over the course of the follow-up campaign to account for the fading of GRB 221009A. Visibilities from each observation block were reduced using a custom pipeline using \texttt{AOFLAGGER} \citep{aoflagger} and \texttt{CASA} \citep{McMullin2007}. Images for the flux, phase and science targets were formed using standard \texttt{CASA} tasks and by deconvolving with the CLEAN algorithm \citep{Hogbom1974, Clark1980, Sault1994}. We used two Taylor terms to account for the high fractional bandwidth (especially at low frequencies) and a Briggs robust value of 0.5 when imaging. Finally, flux densities for GRB 221009A were derived by fitting a point source (i.e., with a source size fixed to the dimensions of the main lobe of the dirty beam) to the science target.

\subsection{ATCA}
We carried out multiple observations of the radio counterpart to GRB 221009A using the Australia Telescope Compact Array (ATCA) under the project codes: CX515 (director's discretionary time), C3374 (PI: G. E. Anderson), C3542 (PI: G. E. Anderson). These observations were carried out using the 5.5/9, 16.7/21.2, 33/35, and 43/45 GHz receiver configurations, with a bandwidth of $2048$\,MHz for each intermediate frequency.

For each observation, we reduced the visibility data using standard procedures in {\sc Miriad} \citep{Sault1995}. We used a combination of manual and automatic radio-frequency interference flagging before calibration. For bandpass calibration, we used PKS B1934$-$638 at 5.5/9\,GHz, while at higher frequencies (16.7/21.2, 33/35 and 43/45\,GHz) we used either B1921$-$293 or B1253$-$055; the spectral shape of B1921$-$293 and B1253$-$055 was accounted for by fitting to first order the measured flux densities of these calibrators at each intermediate frequency for each of the higher frequency observing bands. The flux-density scale was set using B1934$-$638 for all observing frequency bands. For all observations, we used B1923$+$210 to calibrate for the time-variable complex gains.
After calibration, where there was sufficient signal-to-noise, we split the $2048$\,MHz bandwidth into further sub-bands to obtain higher spectral resolution. We then inverted the visibilities and applied the multi-frequency synthesis CLEAN algorithm \citep{Hogbom1974, Clark1980, Sault1994} to the target source field using standard tasks in {\sc Miriad} to obtain our final images.
The flux densities of the radio afterglow candidate were extracted by fitting a point source to the radio source, in the case of a detection, while, in the case of a non-detection, the limits were obtained using the rms sensitivity in the residual image.

\subsection{\textit{e}-MERLIN}

The \textit{enhanced} Multi-Element Remotely Linked Interferometer Network (\textit{e}-MERLIN) is a radio interferometer made up of seven dishes spread across the UK. With a maximum baseline of 217\,km, whilst observing at 5\,GHz, it can resolve angular scales of 0.05''. We observed the position of GRB 221009A with \textit{e}-MERLIN through a combination of rapid response time requests (PI: L. Rhodes, RR14001) and open time proposals (PI: L. Rhodes, CY13003, CY14001 and CY15206) at both L- and C-band. Our L- and C-band observations were centered at 1.51 and 5.08\,GHz, respectively, both with a bandwidth of 512\,MHz. We note that the first two epochs obtained at L-band have previously been published in \citetalias{bright2023}.

All observations were reduced using the \textit{e}-MERLIN pipeline within \textsc{casa} \citep{McMullin2007, 2021ascl.soft09006M}. The pipeline performs preliminary flagging for radio frequency interference and known observatory issues. It then performs two rounds of bandpass calibration and complex gain calibration, using OQ208 and J1905+1943, respectively, along with flux scaling using 3C286. Further flagging of the target field is conducted. We performed interactive cleaning and deconvolution using the \texttt{casa} task \textit{tclean}.


\subsection{LOFAR}
Eight hours of Director’s Discretionary Time with the Low Frequency Array (LOFAR; DDT$20\_003$) were awarded to observe GRB 221009A. The allocated time was split into two observing runs of 4-hours, which took place on 18 and 20 July 2023 at matching local sidereal times. Each observing run was preceded by a 10-minute calibrator scan of 3C295. All observations were conducted in the HBA\_dual\_inner configuration where, in addition to the 22 core stations available, the inner tiles of 14 remote stations were also used. The single-beam observations were centred at 152.05\,MHz with 380 subbands and data were recorded with an integration time of 1 second. Each subband consisted of 64 frequency channels of width 3.051\,kHz.  The data were subsequently averaged to 16 channels of 12.21\,kHz per subband by the observatory during data pre-processing. Both target observations were calibrated for direction independent effects using \textsc{LINC}\footnote{\url{https://linc.readthedocs.io/en/latest/index.html}} with default settings, a pipeline developed by the observatory to correct for various instrumental and ionospheric effects present in interferometric LOFAR data \citep{deGasperin19,vanWeeren16,Williams16}. Due to its relative proximity, Cygnus A was subtracted from the visibilities using the `demixing' step in  \textsc{LINC}. The data were further averaged to 4 channels of 48.82\,kHz per subband and 4 seconds during calibration. The resulting calibrated data were concatenated into groups of 20 subbands and averaged in time to 8 seconds. These data products from both observations were subsequently jointly put through \textsc{ddf-pipeline}\footnote{Second data release version \url{https://github.com/mhardcastle/ ddf-pipeline}. The \texttt{tier1-july2018.cfg} pipeline configuration was used.} for direction-dependent calibration and imaging \citep{Shimwell19,Tasse21}. This resulted in a final image generated using a circular restoring beam of radius 3\,arcseconds and 1.5\,arcseconds pixel resolution. 

\subsection{NOEMA}

The NOrthern Extended Millimetre Array (NOEMA, situated in the southern French Alps) monitored GRB 221009A between October 10\textsuperscript{th} 2022 and April 25\textsuperscript{th} 2023 in the 3, 2 and 1mm bands. Interferometer configurations were medium-extended C and extended A configurations with up to 12 antennas, primary flux calibrators were MWC349 and LKHA101. The data were reduced with the \textsc{clic} and \textsc{mapping} software packages that are part of the \textsc{gildas}\footnote{\url{https://www.iram.fr/IRAMFR/GILDAS}} package. Fluxes and their errors were derived from point-source UV-plane fits to the calibrated interferometric visibilities.

\subsection{uGMRT}
We observed GRB 221009A with the upgraded Giant Metrewave Radio Telescope (uGMRT) in bands 5 (1000—1450 MHz) and 4 (550—900 MHz) under a DDT proposal (ddtC251, PI: P. Chandra). The observations were made at two epochs in both bands, once in January 2023 and then in March 2023. We recorded the data in 2048 frequency channels covering a bandwidth of 400 MHz with an integration time of $\sim10$\,s. We used 3C286 and 3C48 as flux density and bandpass calibrators. J1924+3329 was used as a phase calibrator.

The data were analysed using the \textsc{casa} package \citep[][]{McMullin2007} following the procedure in \citet{2022ApJ...934..186N}. We also performed a few rounds of phase only and one round of amplitude and phase self-calibration to improve the image quality. The final flux densities were obtained by fitting a Gaussian at the GRB position.

\begin{table*}
    \centering
    \begin{tabular}{cccccc}
\hline
Obs Date (MJD)	&	Observing  Frequency 	&	Flux density (mJy)	&	Uncertainty (mJy)	&	Telescope & T-T\textsubscript{0} (days)	\\
\hline	
59866.65	&	15.50	&	7.18	&	0.36	&	AMI	-- LA	 & 5.097 \\
59866.84	&	15.50	&	7.05	&	0.36	&	AMI	-- LA	 & 5.284\\
59867.66	&	15.50	&	6.68	&	0.34	&	AMI	-- LA	 & 6.106\\
59867.84	&	15.50	&	6.84	&	0.35	&	AMI	-- LA	 & 6.283\\
59868.62	&	15.50	&	5.99	&	0.31	&	AMI	-- LA	 & 7.065\\
59869.82	&	15.50	&	5.59	&	0.28	&	AMI	-- LA	 & 8.266\\
...	&	...	&	...	&	...	&	...	 & ... \\
\hline
    \end{tabular}
    \caption{A table of the new radio observations presented in this work. All non-detections are indicated by a `-' in the flux density column followed by the 3$\sigma$ upper limit in the uncertainty column. The full list of radio observations are presented in supplementary material online. }
    \label{tab:radio_obs}
\end{table*}

\section{Results \& Model}\label{sec:model}

There have been several GRB 221009A afterglow modelling efforts which have used a subset of the radio data published to date \citep[including but not limited to][]{laskar2023, oconnor2023, levan2023, 2023MNRAS.524L..78G}. Here, we present the results of our observing campaigns and described out modelling of the radio and X-ray afterglow.

\subsection{Light Curves and SEDs}

\begin{figure*}
    \centering
    \includegraphics[width = 0.8\textwidth]{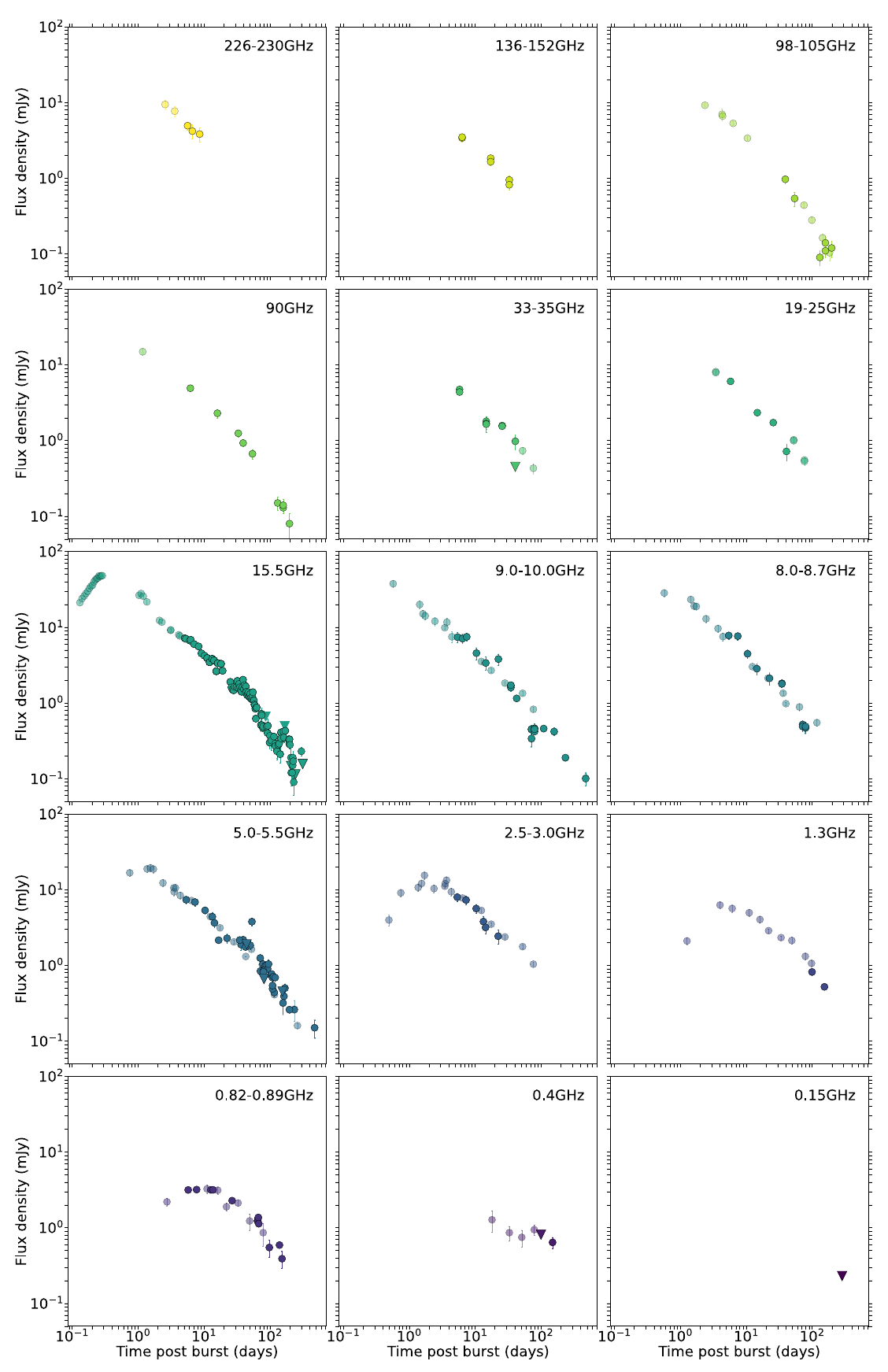}
    \caption{Radio afterglow light curves of GRB 221009A split by observing frequency (or frequency range). Any low-opacity data points are from previously published observations. All observations presented in this paper are shown with solid circles for detections and downwards-facing triangles for 3$\sigma$ upper limits.}
    \label{fig:lightcurves}
\end{figure*}

The radio data presented in this paper spans three orders of magnitude in frequency space, from 0.15\,MHz to 230\,GHz, and lasts out to 475\,days post-burst.  Figure \ref{fig:lightcurves} shows the radio afterglow light curves split by observing frequency. Symbols with lower opacity denote all previously published data whereas the solid symbols mark data presented in this paper. We include all previous and newly published data to extract the clearest scenario of the afterglow. 

Above 19\,GHz, the afterglow is decaying at all times, with observations obtained between 1 and 200\,days post-burst (the top two rows of Figure \ref{fig:lightcurves}). The light curves between 90 and 105\,GHz in Figure \ref{fig:lightcurves} show that the decay rate slowly steepens with time like a very smooth broken power law. Below 16\,GHz, we observe the light curve peak in almost each observing band, except at 9--10 and 0.4\,GHz since we were not observing early enough at those frequencies. \citetalias{bright2023} interpreted this peak as emanating from the reverse shock, which we are tracking from 17.7 to below 1\,GHz. The data between 1.3 and 3\,GHz also show a second, distinct bump at around 50\,days. In addition to the early peaks caught at 5 and 15.5\,GHz, we also see evidence of further bumps during the decay phase. It is possible that the additional bumps originate from different spectral components.

\begin{figure*}
    \centering
    \includegraphics[width = \textwidth]{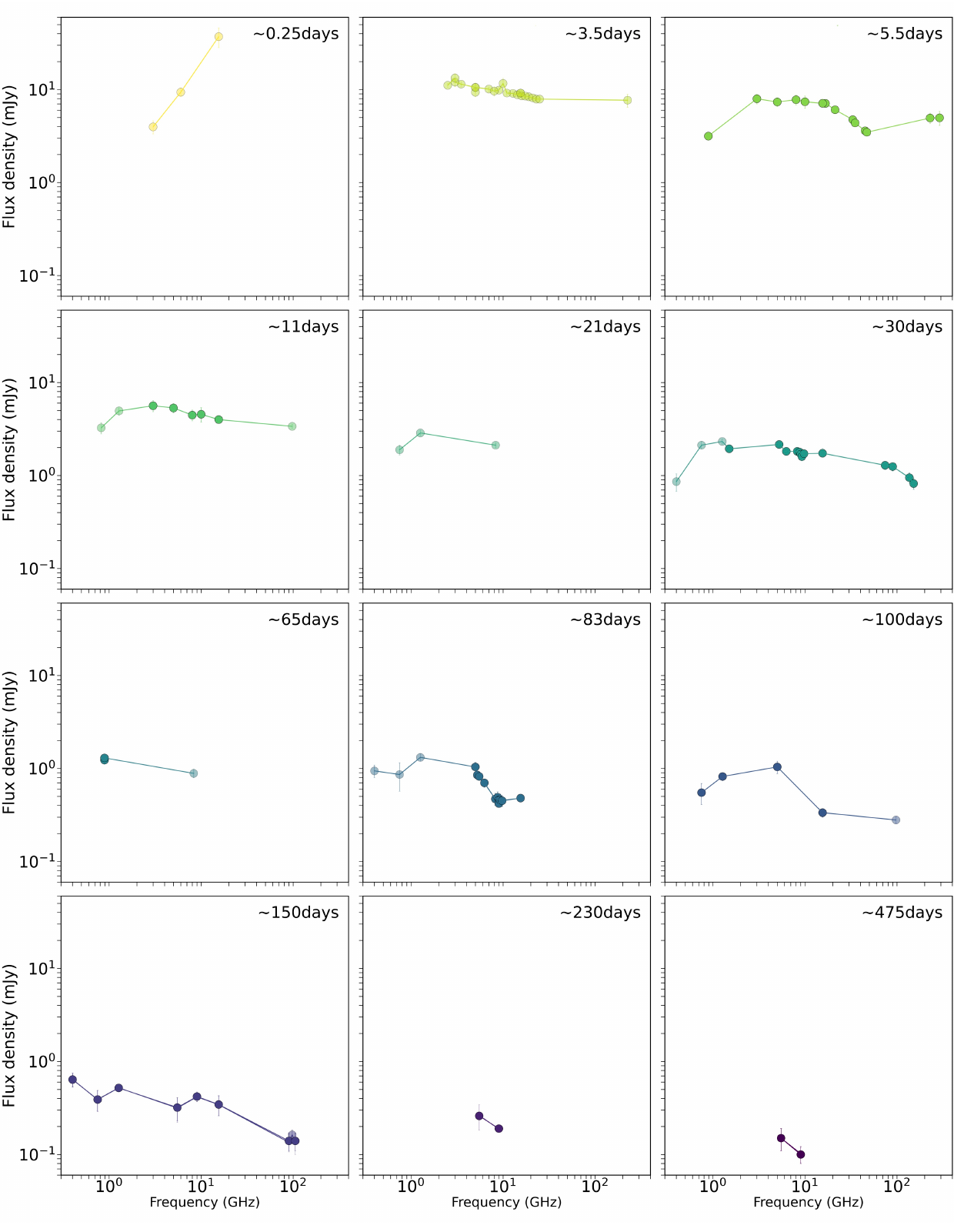}
    \caption{Broadband radio SEDs for GRB 221009A as a function of time. As in Figure \ref{fig:lightcurves}, low-opacity data points denote previously published data, while solid points are observations presented in this paper. Because epochs have been chosen to demonstrate the spectral evolution, we note that not all data presented in Figure \ref{fig:lightcurves} are also shown here.}
    \label{fig:seds}
\end{figure*}

Figure \ref{fig:seds} shows the broadband radio spectral energy distributions (SEDs) throughout our campaign. For the first 30\,days, a low-frequency turnover is visible and the below-turnover spectral index is consistent with $\beta \sim 5/2$ below the turnover. Above the turnover, we find a flat spectrum extending to the highest frequencies ($\sim200$\,GHz). A flat spectrum is inconsistent with optically thin synchrotron emission from a single component and so provides further evidence of multiple spectral components, similar to GRB\,130427A \citep{perley2014}. Only after 150\,days post-burst does the spectrum steepen with typical optically thin spectral indices ($\beta \sim -0.5$ to $-1$), more consistent with that from the late-time X-ray data \citep{williams2023}. \citet{williams2023} performed a joint fit to the UV, X-ray and gamma-ray data, which shows that the high-energy spectra can be described by either a single power law or a broken power law where the break, interpreted as the synchrotron cooling break $\nu_{\textrm{c}}$, sits in the XRT band. The broken power law is favoured but the fits are only performed on data up to one day post-burst whereas the X-ray light curve itself extends out to 200\,days post-burst. 

\subsection{Modelling}

Here, we build on previous modelling efforts by combining our new observations from AMI--LA, ATA, ATCA, ASKAP, \textit{e}-MERLIN, LOFAR, NOEMA and uGMRT with radio data available in the literature. We also include the full \textit{Swift}-XRT light curve \citep[in flux densities at 10\,keV;][]{williams2023}. We do not include any optical or other high-energy data in our modelling work as there are too many contaminating components in these bands. At optical frequencies, there is significant extinction \citep{2023ApJ...946L..30T,2023MNRAS.521.1590V} both from the Milky Way and the host galaxy, plus a contribution from the associated supernova. Above keV energies, there is an increasing contribution from the additional VHE component whose origin and emission mechanism is still debated \citep{ 2023ApJ...946L..27A,lhaaso2023,INTEGRAL2024}. 

We consider models that use either two or three synchrotron spectral components that can evolve independently in time to explain the behaviour shown in the light curves (Figure \ref{fig:lightcurves}) and SEDs (Figure \ref{fig:seds}). 
Each synchrotron spectrum is constructed of four power-law slopes divided by three frequency breaks: the synchrotron self-absorption break ($\nu_{\textrm{sa}}$), the characteristic or minimum electron energy break ($\nu_{\textrm{m}}$), and the cooling break ($\nu_{\textrm{c}}$, above which radiative cooling is important). The peak of the spectrum, $F_{\nu, \textrm{max}}$, is at whichever frequency break of $\nu_{\textrm{sa}}$ or $\nu_{\textrm{m}}$ is higher. The spectral index of each branch depends on the order of the frequency breaks. In the regime where $\nu_{\textrm{sa}} < \nu_{\textrm{m}} < \nu_{\textrm{c}}$, the spectral indices are $F_{\nu<\nu_{\textrm{sa}}} \propto \nu^{2}$, $F_{\nu_{\textrm{sa}} < \nu <\nu_{\textrm{m}}} \propto \nu^{1/3}$, $F_{\nu_{\textrm{m}} < \nu<\nu_{\textrm{c}}} \propto \nu^{(1-p)/2}$ and $F_{\nu_{\textrm{c}} < \nu} \propto \nu^{-p/2}$, where $p$ is the electron energy distribution index and is typically expected to be between 2 and 3  \citep[although values slightly below 2 and above 3 have been reported, ][]{2000ApJ...542..235K, 2001MNRAS.328..393A,2013ApJ...771...54S}. In the regime where $\nu_{\textrm{m}} < \nu_{\textrm{sa}} < \nu_{\textrm{c}}$, the spectral indices are $F_{\nu<\nu_{\textrm{m}}} \propto \nu^{2}$, $F_{\nu_{\textrm{m}} < \nu <\nu_{\textrm{sa}}} \propto \nu^{5/2}$, $F_{\nu_{\textrm{sa}} < \nu<\nu_{\textrm{c}}} \propto \nu^{(1-p)/2}$ and $F_{\nu_{\textrm{c}} < \nu} \propto \nu^{-p/2}$. As the jet expands and evolves, the spectral breaks are expected to change as a power-law function of time, which depends on the jet dynamics and the density profile through which the jet is propagating, $\rho \propto r^{-k}$, where $k = 0$ for a homogeneous medium and $k = 2$ represents a stellar wind \citep{granot2002, 2014PASA...31....8G}.

We use \textsc{emcee} to fit our respective models to the data \citep{Foreman-Mackey2013}. Each model uses 40 walkers and runs for at least 70000 steps or until convergence. All priors are uniform, and the only priors with fixed bounds were $p \in [1.5,3.5]$ to help rule out unphysical solutions. 
The best fit value for each parameter is the 50\textsuperscript{th} percentile post burn in of the posterior distribution, and the 84\textsuperscript{th} and 16\textsuperscript{th} percentiles are quoted as the upper and lower uncertainties, respectively.

\subsubsection{Two-Component Model}

First, we fit the data with two separate synchrotron spectra. The first is the reverse shock identified in \citetalias{bright2023}, we find that the peak of the synchrotron spectrum is produced by $\nu_{\textrm{sa}}$ and fit for the normalisation and evolution of the spectrum as well as $p$. The second component is a forward shock that appears to dominate the optical and X-rays \citep[e.g.][]{williams2023, Fulton2023, Shrestha2023}, and also the late-time radio emission. Here we allow both $\nu_{\textrm{sa}}$ and $\nu_{\textrm{m}}$ to vary freely. We fit for the normalisation and evolution of $F_{\nu, \textrm{max}}$, $\nu_{\textrm{sa}}$ and $\nu_{\textrm{m}}$ as well as $p$. The resulting model parameters are provided in Table \ref{tab:model_results_2comp}.

We find that the two-component model cannot reproduce the flat spectrum observed shown in Figure \ref{fig:seds}, the posterior distribution of $p$ for the reverse shock always ends up at the lower bound of the prior with values for $p$ below 1.5 or even below 1, and such a low value is unphysical and so we no longer consider this scenario.

\begin{table}
    \centering
    \begin{tabular}{c c}
    \hline
    \hline
        Parameter & Value  \\
        \hline
                \hline
        \multicolumn{2}{c}{Reverse shock}\\
        \hline
        F$_{\nu, \textrm{max}}$\,(mJy)[1\,day] & 24.0$\pm$0.8\\
        $\nu_{\textrm{sa}}$\,(GHz)[1\,day] & 6.3$\pm$0.1\\
        $\alpha_{\textrm{F}_{\nu, \textrm{max}}}$ & -0.84$\pm$0.02\\
        $\alpha_{\textrm{sa}}$ &-0.957$\pm$0.008\\
        p& $<$1.5\\
        \hline
        \multicolumn{2}{c}{Forward shock}\\
        \hline
        F$_{\nu, \textrm{max}}$\,(mJy)[6.5\,days]  & 3.10$\pm$0.06\\
        $\log(\nu_{\textrm{sa}})$\,(GHz)[6.5\,days]   & -0.5$\pm$0.1\\
        $\log(\nu_{\textrm{m}})$\,(GHz)[6.5\,days]  &  2.20$\pm$0.04\\
        $\alpha_{\textrm{F}_{\nu, \textrm{max}}}$ & -0.63$\pm$0.02\\
        $\alpha_{\textrm{m}}$ & -1.67$\pm$0.03\\
        $\alpha_{\textrm{sa}}$ & -0.11$\pm$0.07\\
        p & 2.32$\pm$0.03\\
        \hline
    \end{tabular}
    \caption{The parameter values (50\textsuperscript{th} percentile) and their associated uncertainties (18\textsuperscript{th} and 64\textsuperscript{th} percentiles) derived for our best-fit two-component model. Any $\alpha$ parameter refers to the temporal power law index of the parameter written in the subscript, as described in Section \ref{sec:model}. For the reverse, forward and extra shock component, F$_{\nu, \textrm{max}}$ and $\nu_{\textrm{sa}}$ are normalised to 1\,day and 6.5\,days, respectively. For each shock, p is the value of the electron energy spectral index. }
    \label{tab:model_results_2comp}
\end{table}

\subsubsection{Three-Component Model}

Given the issues with a two-component model, we include a third component to alleviate the shallow value of \textit{p} which was needed in the two-component model to explain the flat spectrum that is present during the first $\sim150$\,days (Figure~\ref{fig:seds}) and the additional bumps in the 5 and 15.5\,GHz light curves around 5--10\,days post-burst (see Figure~\ref{fig:lightcurves}).  

To best explore the parameter space of the third component, first, we test both $\nu_{\textrm{sa}}$ and $\nu_{\textrm{m}}$ as the peak frequency of the third component and find that $\nu_{\textrm{sa}}$ provides a better fit. Then we consider two different iterations of this extra shock with differing degrees of freedom, which are summarised in Table \ref{tab:extra_shock}, in addition to the two shock components described in the previous section. In both iterations of our three-component model, the peak flux density of each of the three components follows a smoothly broken power law \citep{2020MNRAS.496.3326R}: 

\begin{equation}
    F_{\nu} = F_{\nu, \textrm{max}} \left(0.5\left(\frac{t}{t_{b}} \right)^{-\alpha_{1}s} +0.5\left(\frac{t}{t_{b}} \right)^{-\alpha_{2}s} \right)^{-\frac{1}{s}}
    \label{eq:brkn_pwl}
\end{equation}

\noindent where $F_{\nu, \textrm{max}}$ is the flux density at the break time $t_{b}$, $\alpha_{1}$ and $\alpha_{2}$ are the power-law indices, and $s$ is the smoothing parameter which we set to be 0.5. In Model 1, the synchrotron self-absorption break follows a single power law: $\nu_{\textrm{sa}} = \nu_{\textrm{sa}, 0} t^{\alpha_{\textrm{sa}}}$ where $\nu_{\textrm{sa}, 0}$ is the location of the self-absorption break at 1\,day post-burst. We set $\alpha_{F_{\nu, \textrm{max}, 1}} = 3$ and both $\alpha_{F_{\nu, \textrm{max}, 2}}$ (defined in Table \ref{tab:extra_shock} as $\alpha$) and $\alpha_{\textrm{sa}}$ can vary freely. We invoke a $\alpha_{F_{\nu, \textrm{max}, 1}} = 3$ as done in \citet{2005ApJ...626..966P} which is used in the regime where a blastwave that is initially off-axis has undergone significant deceleration and so the radiation begins to enter the observers' line of sight. In the paper, they do not consider the self-absorption break, but we find it fits well within the constraints of our work. \citet{2020ApJ...896..166R} also consider off-axis afterglows from a numerical perspective and find steeper rise rates for `far off-axis events'. We choose to be more conservative and use \citet{2005ApJ...626..966P} value.

In model 2, both the peak flux density and $\nu_{\textrm{sa}}$ are both described with broken power laws where all the indices are fit for but the break time is the same. A full summary of our models to explain the extra forward shock is shown in Table \ref{tab:extra_shock}.

\begin{figure*}

 \begin{subfigure}[t]{0.42\textwidth}
        \centering
        \includegraphics[width=\linewidth]{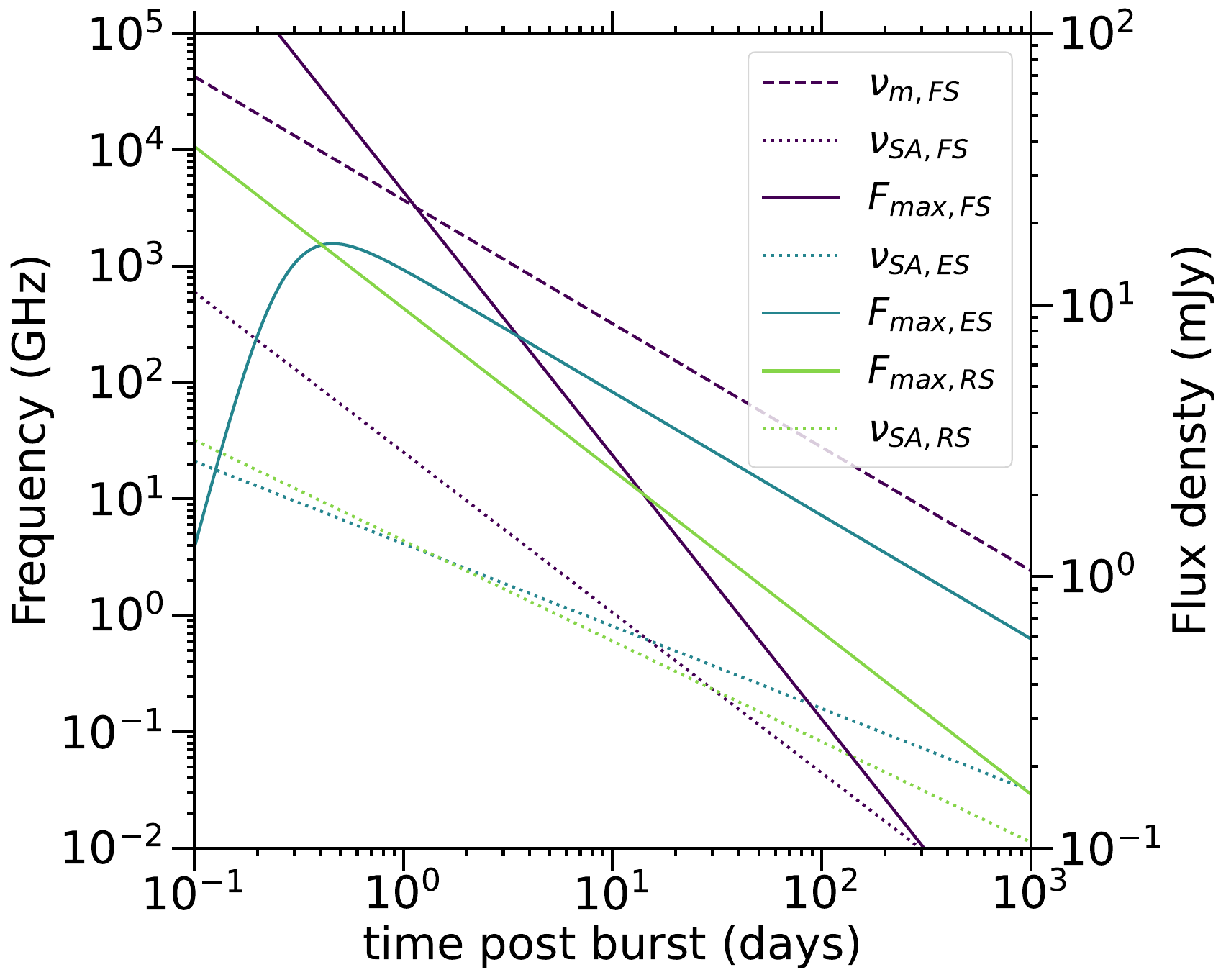} 
        \caption{Model 1}\label{fig:model3}
    \end{subfigure}
    \begin{subfigure}[t]{0.4\textwidth}
    \centering
        \includegraphics[width=\linewidth]{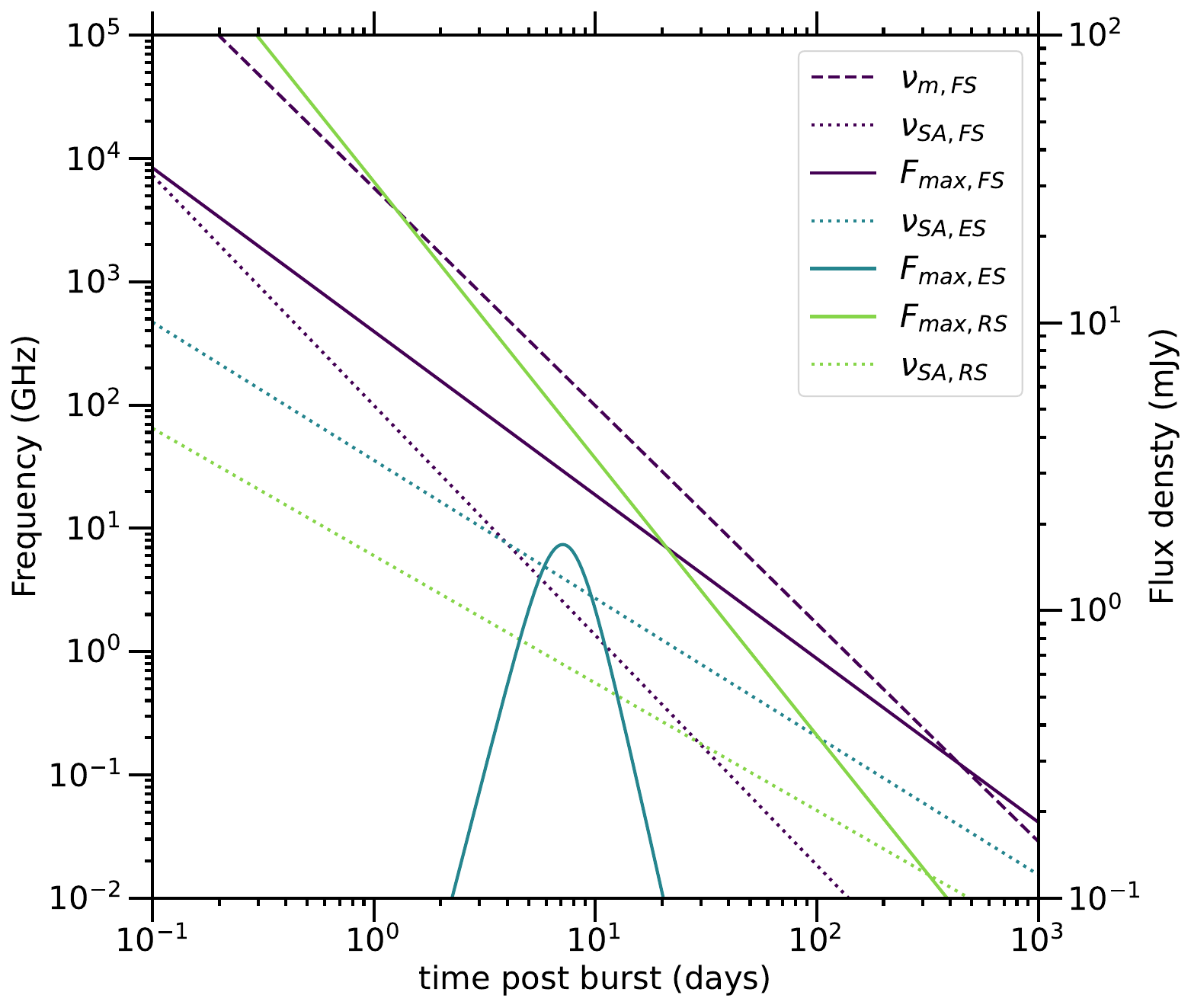} 
        \caption{Model 2} \label{fig:model4}
    \end{subfigure}

    \caption{Evolution of the break frequencies and peak flux for the three-component model. Each panel corresponds to a different iteration of our model as described in Section \ref{sec:model} and Table \ref{tab:extra_shock}. For each iteration, we show only the average value (50\textsuperscript{th} percentile value) of the posterior distribution for clarity. The lefthand vertical axis of each plot corresponds to the evolution of the frequency breaks (dotted and dashed lines for $\nu_{\textrm{sa}}$ and $\nu_{\textrm{m}}$, respectively). The righthand vertical axis shows the evolution of the peak flux (solid lines) of each shock component.}
    \label{fig:diff_models}
\end{figure*}

\begin{figure*}
    \centering
    \includegraphics[width = 0.8\textwidth]{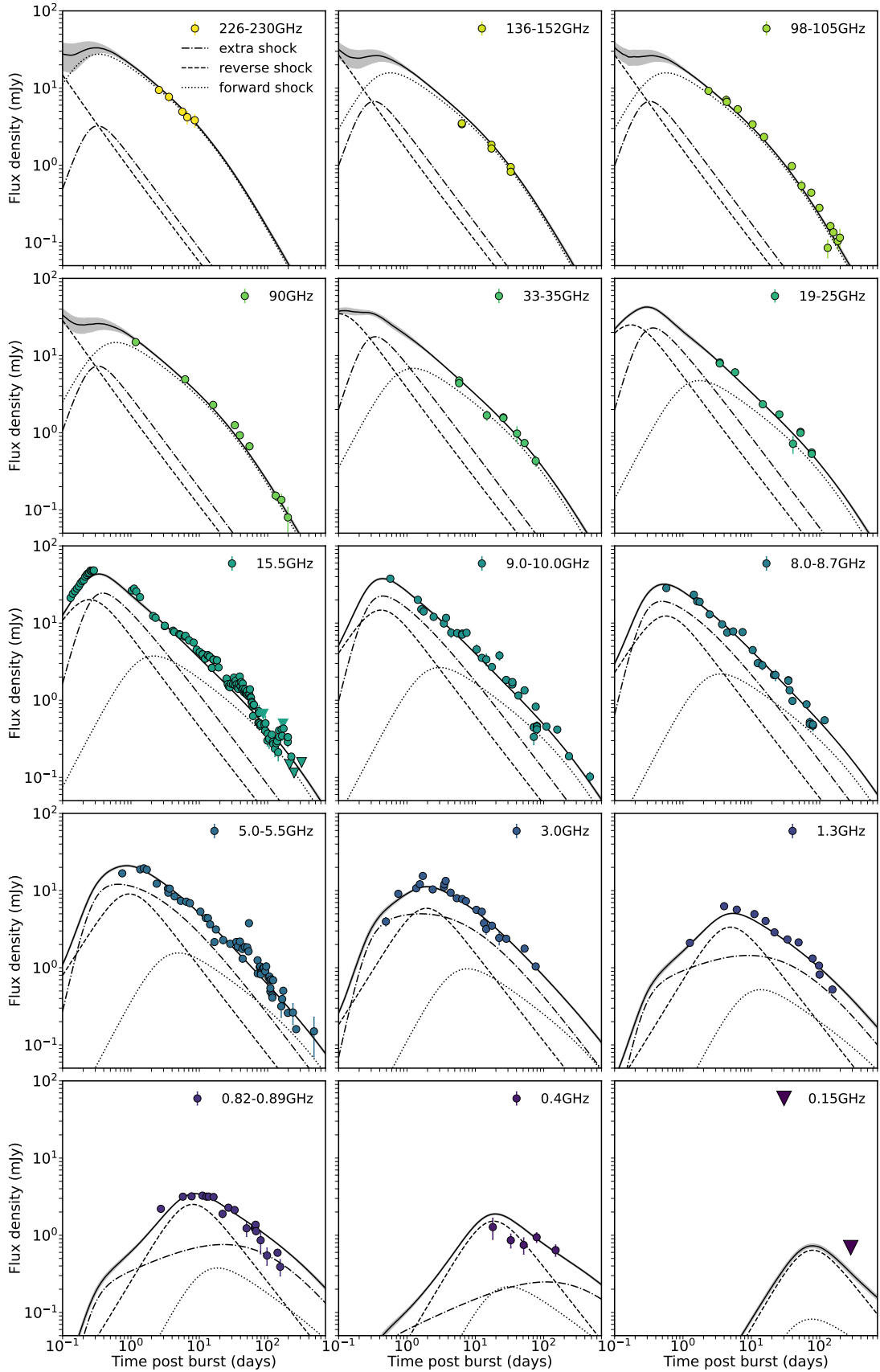}
    \caption{The multi-frequency radio light curves for GRB 221009A overlaid with our best-fit three-component model (model 1).}
    \label{fig:lightcurve_model}
\end{figure*}

\begin{figure*}
    \centering
    \includegraphics[width = \textwidth]{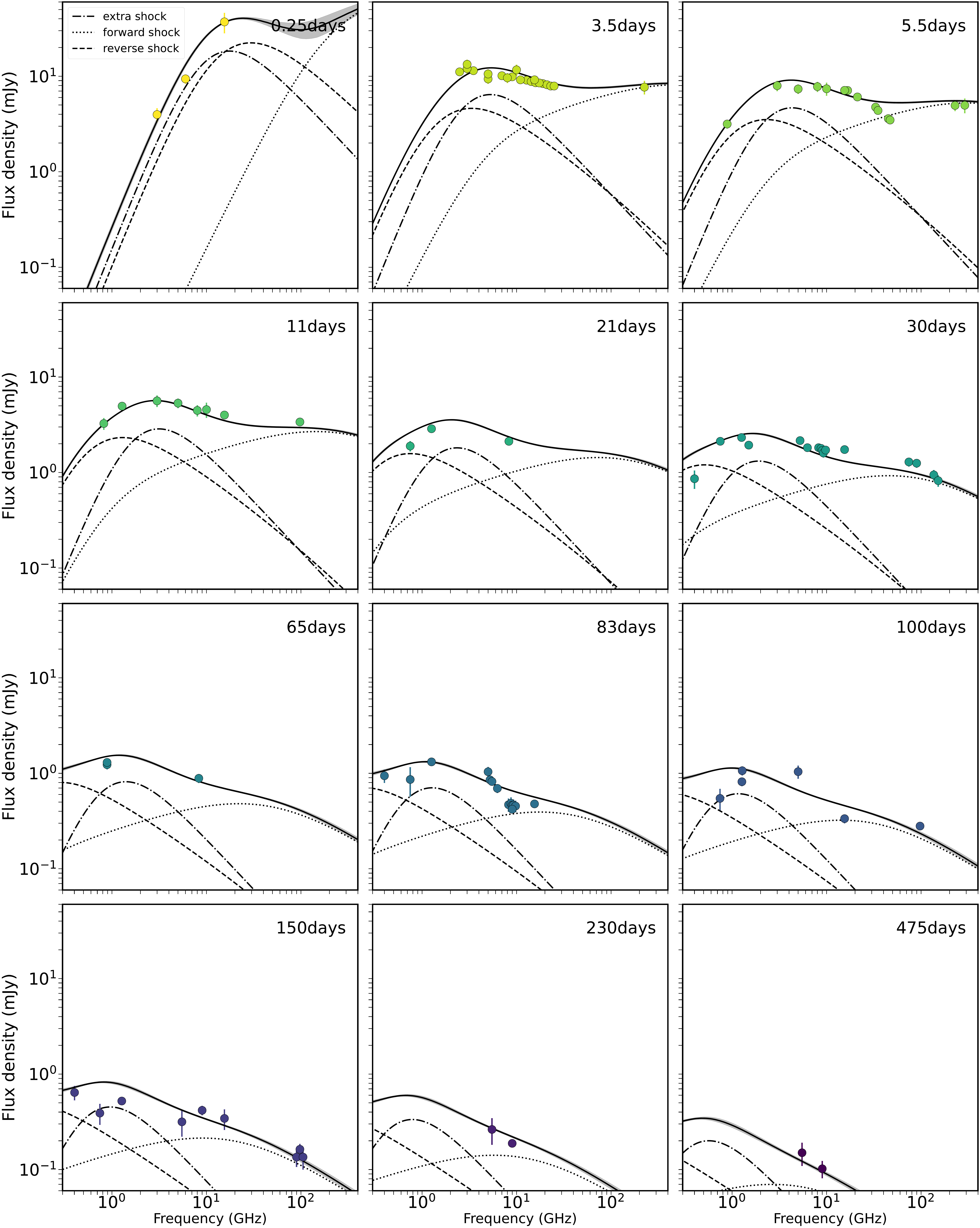}
    \caption{Broadband radio SEDs for GRB 221009A as a function of time with our best-fit three-component model (model 1) overlaid.}
    \label{fig:sed_model}
\end{figure*}

Figure \ref{fig:diff_models} shows the results of the different iterations of our models. Unfortunately, not one of our models provides a perfect fit to the data, this may be due to a combination of unknown systematic uncertainties, and perhaps more importantly, this exquisite data set is showing evidence of more complicated physics and emission mechanisms that cannot be accounted for by the basic synchrotron models. As a result, we find quoting Bayesian evidence values inappropriate. However, we do find that \textit{model 1} provides the best fit. This is because our posterior distributions for all values of $p$ sit between 2 and 3 and do not require such uncomfortably large temporal index values. We present the parameters of this fit in Table \ref{tab:model_results}. Figures \ref{fig:lightcurve_model} and \ref{fig:sed_model} show our best-fitting model overlaid on the light curves and SEDs. Figure \ref{fig:lightcurve_model} shows that our model describes the long term evolution at all frequencies well. However, it cannot replicate the bumps and wiggles observed at 15.5, 5 and 0.4\,GHz, despite that being one of the motivations for the three component model. Furthermore, it marginally over predicts the late time 0.8\,GHz flux. Figure \ref{fig:sed_model} demonstrates that the superposition of multiple components recreates the high frequency emission accurately and describes well the flat spectrum and broad turnover at earlier times post-burst. On the other hand, we find that it tends to place the $\nu_{\textrm{sa}}$ much lower than the observed position.

\begin{table}
    \centering
    \begin{tabular}{ccccc}
    \hline
        Model $\#$ & $\alpha_{F_{\nu, \textrm{max}, 1}}$& $\alpha_{F_{\nu, \textrm{max}, 2}}$ &$\alpha_{\nu_{\textrm{sa}}, 1}$ &$\alpha_{\nu_{\textrm{sa}}, 2}$\\
        \hline
        1 & $3$&$\alpha$ & $\alpha_{\textrm{sa}}$ & -\\
        2 & $\alpha_{1}$&$\alpha_{2}$ & $\alpha_{\textrm{sa}, 1}$& $\alpha_{\textrm{sa}, 2}$\\
        \hline
    \end{tabular}
    \caption{Summary of the different iterations of the three-component model which explore the possible evolution of the third shock component. Each $\alpha$ corresponds to a temporal index of the subscripted value, e.g., $\alpha_{F_{\nu, \textrm{max}, 1}}$ corresponds to the first temporal index used to describe the behaviour of $F_{\nu, \textrm{max}}$. We find that model 1, combined with a forward and reverse shock describes the data best. The model is shown compared to the data in Figures \ref{fig:lightcurve_model} and \ref{fig:sed_model}. The best fit parameters are shown in Table \ref{tab:model_results}.}
    \label{tab:extra_shock}
\end{table}

\begin{figure}
    \centering
    \includegraphics[width = \columnwidth]{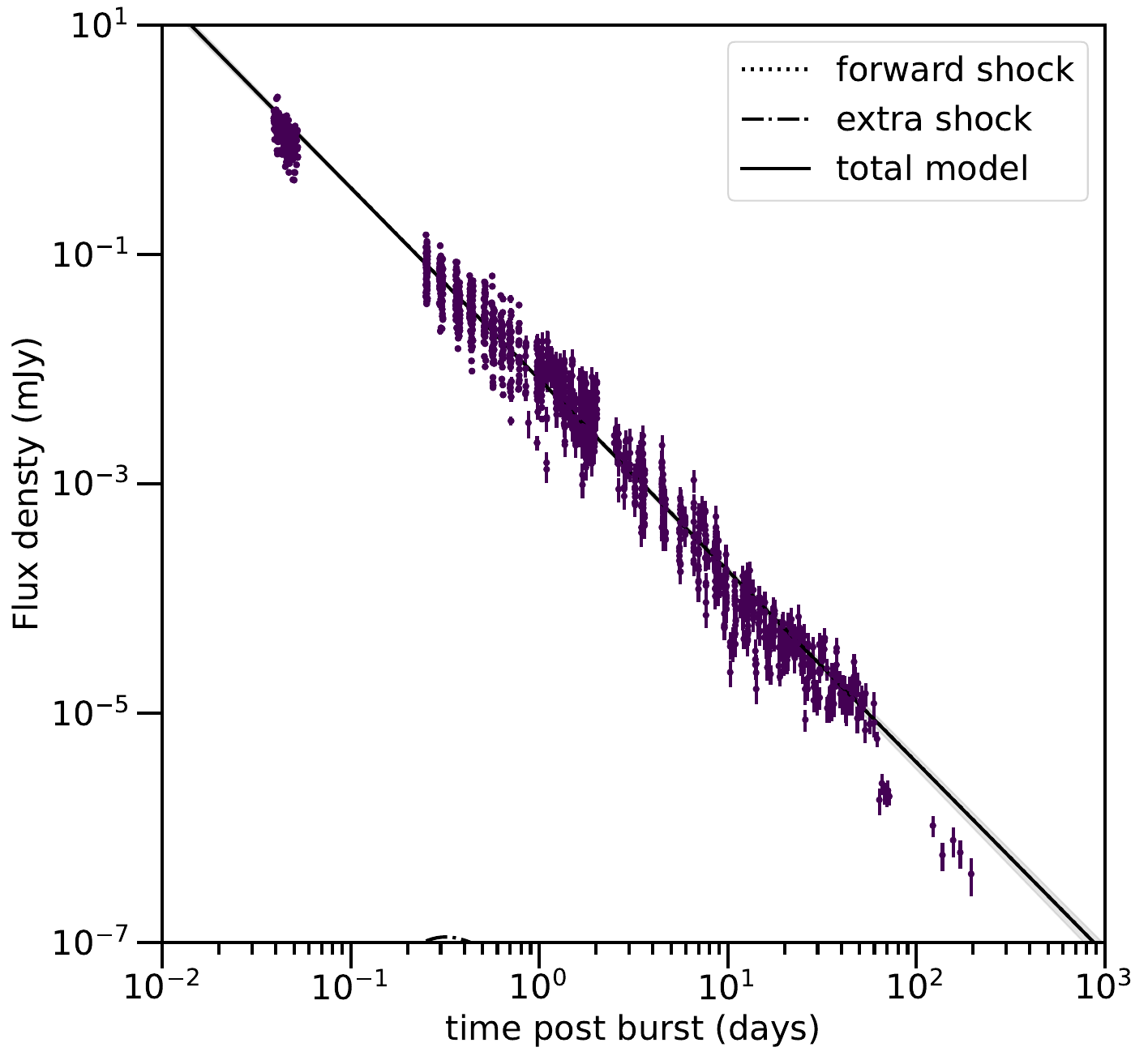}
    \caption{The X-ray light curve for GRB 221009A at 10\,keV, overlaid with our best-fit afterglow model.}
    \label{fig:xray_model}
\end{figure}

\begin{figure}
    \centering
    \includegraphics[width = \columnwidth]{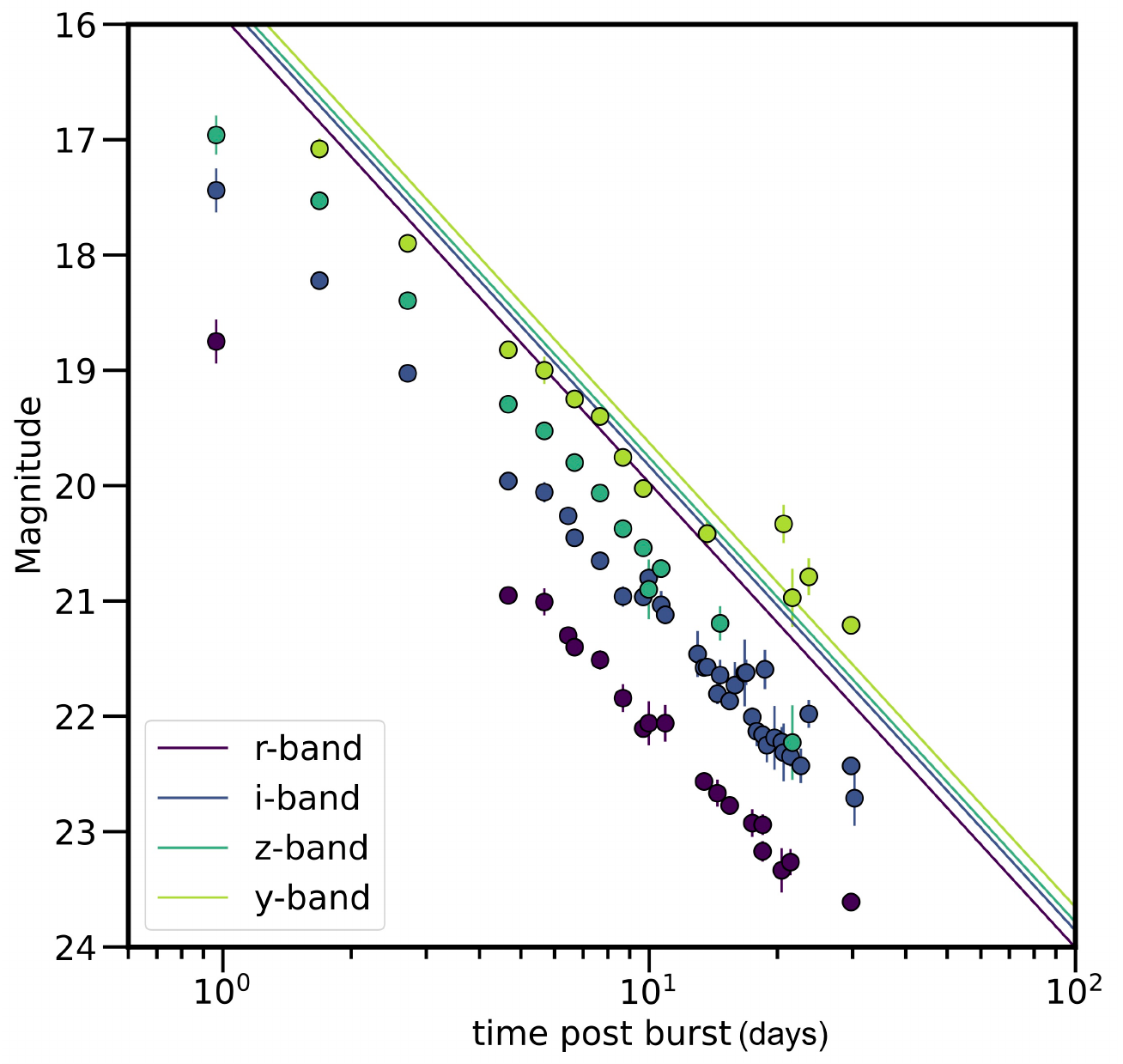}
    \caption{Optical light curves of GRB 221009A from \citet{Fulton2023}, overlaid with our best-fit afterglow model. While we do not fit our model to the optical data due to the large and mostly unconstrained extinction contribution as well as the supernova \citep{2023ApJ...948L..12K, 2024NatAs.tmp...65B}, our model reproduces the decay rate of the optical data well. It is clear that significant extinction, 2 magnitudes in the r-band, is needed to get the correct normalisation of our model with respect to the data. }
    \label{fig:optical_model}
\end{figure}

\begin{table}
    \centering
    \begin{tabular}{c c}
    \hline
    \hline
        Parameter & Value  \\
        \hline
                \hline
        \multicolumn{2}{c}{Reverse shock}\\
        \hline
        F$_{\nu, \textrm{max}}$\,(mJy)[1\,day] & $9.6^{+1.6}_{-1.5}$\\
        $\nu_{\textrm{sa}}$\,(GHz)[1\,day] & $4.4^{+0.2}_{-0.3}$\\
        $\alpha_{\textrm{F}_{\nu, \textrm{max}}}$ & -0.59$\pm$0.05\\
        $\alpha_{\textrm{sa}}$ &-0.86$\pm$0.03\\
        p& $2.2^{+0.4}_{-0.3}$\\
        
        \hline
        \multicolumn{2}{c}{Forward shock}\\
        \hline
        F$_{\nu, \textrm{max}}$\,(mJy)[6.5\,days]  & 4.2$\pm$0.2\\
        $\log(\nu_{\textrm{sa}})$\,(GHz)[6.5\,days]   & 0.3$\pm$0.2\\
        $\log(\nu_{\textrm{m}})$\,(GHz)[6.5\,days]  &  2.71$\pm$0.08\\
        $\alpha_{\textrm{F}_{\nu, \textrm{max}}}$ & -0.97$\pm$0.03\\
        $\alpha_{\textrm{m}}$ & -1.06$\pm$0.06\\
        $\alpha_{\textrm{sa}}$ & $-1.4^{+0.2}_{-0.1}$\\
        p & 2.32$\pm$0.03\\

        \hline
        \multicolumn{2}{c}{Extra shock}\\
        \hline
        F$_{\nu, \textrm{max}}$\,(mJy)[t$_{\textrm{dec}}$\,days] & 17$\pm$2\\
        $\nu_{\textrm{sa}}$\,(GHz)[t$_{\textrm{dec}}$\,days] & $1.03^{+0.05}_{-0.04}$\\
        $\alpha$ & -0.71$\pm$0.02\\
        $\alpha_{\textrm{sa}}$ & $-0.46^{+0.03}_{-0.04}$\\
        t$_{\textrm{dec}}$\,(days) & 0.27$\pm$0.02\\
        p & 3.1$\pm$0.3\\
        \hline
    \end{tabular}
    \caption{The parameter values (50\textsuperscript{th} percentile) and their associated uncertainties (18\textsuperscript{th} and 64\textsuperscript{th} percentiles) derived for our best-fit three-component model (model 1). Any $\alpha$ parameter refers to the temporal power law index of the parameter written in the subscript, as described in Section \ref{sec:model} and Table \ref{tab:extra_shock}. For the reverse, forward and extra shock component, F$_{\nu, \textrm{max}}$ and $\nu_{\textrm{sa}}$ are normalised to 1\,day, 6.5\,days and t$_{\textrm{dec}}$, respectively, where t$_{\textrm{dec}}$ is a parameter we fitted for. For each shock, p is the value of the electron energy spectral index. }
    \label{tab:model_results}
\end{table}

\section{Discussion}\label{sec:discussion}

Here we discuss the implications of our best-fitting three-component model and place them in context of other detailed radio studies of GRBs.

\subsection{Reverse shock}

The dashed lines in Figures \ref{fig:lightcurve_model} and \ref{fig:sed_model} show the contribution of the reverse shock from our model. \citetalias{bright2023} used radio observations in the first five days post-burst to measure the evolution of F\textsubscript{max} and $\nu_{\textrm{sa}}$ with time. They found that $F_{\nu, \textrm{max}} \propto t^{-0.70\pm0.02}$ and $\nu_{\textrm{sa}} \propto t^{-1.08\pm0.04}$, and concluded that the evolution of the spectral peak was too slow to match theoretical predictions and most likely a superposition of multiple emitting regions. When considering the full radio data set, we find a different, even slower reverse shock evolution: F\textsubscript{max} $\propto t^{-0.59\pm0.05}$ and $\nu_{\textrm{sa}} \propto t^{-0.86\pm0.03}$, and that multiple shocks are contributing to the early 15.5\,GHz observation. We find that the slow evolution of the reverse shock means that it contributes significantly to the low-frequency emission at all times. 

To contextualise these findings, we compare our results to both thin and thick reverse shock models summarised in \citet{vanderhorst2014}. The distinction between thin and thick shell models refers to the depth and velocity spread of the shell that the shock is moving through. The reverse shock emission is produced as it propagates back through the shell at the front of the jet. In a thick shell scenario, the velocity spread of the ejected material is large enough such that the shock can accelerate to become relativistic, and the resulting light curves depend on the circumburst environment profile, as does the forward shock. In the thin shell scenario, the reverse shock remains Newtonian, and reverse shock light curves are dependent on the deceleration profile of the jet \citep{1995ApJ...455L.143S,1999MNRAS.306L..39M}. With the results of our model, we cannot recreate our observations with physically realistic parameter values for either a thick or thin shell reverse shock model. We find that the reverse shock evolution that we measure is too slow compared to analytical models such as those in \citet{vanderhorst2014}.

Compared to the number of detailed forward shock studies, there are very few GRBs where the reverse shock is observed in sufficient detail to confidently examine certain reverse shock models. GRBs 130427A, 190114C and 190829A are the three most well-studied GRBs with bright reverse shock components \citep[they also happen to all have -- at least tentative -- very high energy components like GRB 221009A;][]{2014Sci...343...42A,2019Natur.575..459M,2021Sci...372.1081H}. The reverse shock component from GRB 190114C appears to match with theoretical models for reasonable physical parameters \citep{2019ApJ...878L..26L}. However, GRBs 130427A and 190829A could not be explained by analytical reverse shock models \citep{vanderhorst2014,2022ApJ...931L..19S}. In the case of GRB 190829A, the best fit came from assuming a rapid decay in the magnetic field strength post-shock crossing \citep{2022ApJ...931L..19S}. It is possible that GRB 221009A requires a similarly complex model to explain the observed behaviour but that is beyond the scope of this work.

\subsection{Forward shock}

The dotted lines in Figures \ref{fig:lightcurve_model} and \ref{fig:sed_model} denote the contribution from the forward shock. The forward shock component of our model dominates all of the high-frequency light curves (above 33\,GHz) at all times. Moving to lower observing frequencies the forward shock contributes less, and below 10\,GHz the forward shock component is always subdominant. At X-ray energies (Figure \ref{fig:xray_model}), the emission is always dominated by the forward shock component (the dotted line). Given how well our model fits the X-ray data, the cooling break $\nu_{\textrm{c}}$ seems to be situated above the X-ray regime throughout the observations. Although we do not fit our model to the optical data, we have overlaid our model onto the optical data from \citet{Fulton2023} in Figure \ref{fig:optical_model}. The decay rate of our model matches that of the data except for the late time y-band data, which \citet{Fulton2023} suggested was due to a supernova component. Figure \ref{fig:optical_model} reinforces that there is significant extinction affecting the optical emission from GRB 221009A \citep{Fulton2023, levan2023, 2023ApJ...948L..12K, 2023ApJ...946L..30T}. We find that nearly 2 magnitudes of extinction are needed in the r-band, decreasing to $\sim0.1{-}0.2$ magnitudes in the y-band.  

Traditional forward shock spectral models take the three frequency breaks and the peak flux density and calculate four afterglow parameters: the total kinetic energy, the circumburst density, the fraction of kinetic energy that goes into the electrons and magnetic fields \citep[][]{sari1998, Chevalier1999}. From there, if a jet break is detected (an achromatic break in the light curves), the opening angle of the jet can be calculated \citep{1999ApJ...519L..17S}. 
For GRB 221009A, we cannot calculate these parameters for two main reasons. The first is that whilst we are able to track the evolution of $F_{\nu, \textrm{max}}$, $\nu_{\textrm{m}}$ and $\nu_{\textrm{sa}}$ for the forward shock, with the data we use in this work we are unable to localise $\nu_{\textrm{c}}$ since it appears to be above the X-ray band \citep{williams2023}, and $\nu_{\textrm{c}}$ is needed to break the degeneracy between the different afterglow parameters. Secondly, to calculate the afterglow parameters, the observed evolution must match the model's prediction. Otherwise, the afterglow parameters derived at each time step will have different values. 

Our model finds that $\nu_{\textrm{m}} \propto t^{-1.06\pm0.06}$, whereas theoretically it is expected that  $\nu_{\textrm{m}} \propto t^{-1.5}$ independent of circumburst environment density profile, strongly in disagreement with our findings. We also find that $F_{\nu, \textrm{peak}}$ and $\nu_{\textrm{sa}}$ do not evolve in agreement with expectations from the standard afterglow model, instead we find that $F_{\nu, \textrm{peak}} \propto t^{-0.97\pm0.03}$ and $\nu_{\textrm{sa}} \propto t^{-1.4^{+0.2}_{-0.1}}$ (we note that the temporal index for $\nu_{\textrm{sa}}$ is pushing up on the bounds set for the priors in the \textsc{emcee} fit). Comparatively, for a stellar wind ($k = 2$) and homogeneous ($k = 0$) environment, $F_{\nu, \textrm{peak}}$ is expected to evolve as $t^{-0.5}$ and $t^{0}$ \citep{granot2002}, respectively, which is far slower than what we observe. The expected evolution of the synchrotron self-absorption break is also dependent on the circumburst environment's density profile: with $t^{0}$ and $t^{-0.6}$ for $k = 0$ and $k = 2$, respectively, again the temporal indices are too slow to match our model. 

Using the relations from table 5 in \citet{vanderhorst2014}, we can derive individual circumburst density profiles from the evolution of both $\nu_{\textrm{sa}}$ and $\textrm{F}_{\nu, \textrm{max}}$. We find that $\nu_{\textrm{sa}} \propto t^{-1.4^{+0.2}_{-0.1}}$ and $\textrm{F}_{\nu, \textrm{max}} \propto t^{-0.97\pm0.03}$ corresponds to $k = 2.8^{+0.10}_{-0.06}$ and $k = 2.64\pm0.03$, respectively. Both the evolution of $F_{\nu, \textrm{max}}$ and $\nu_{\textrm{sa}}$ strongly favour a steeper circumburst density profile over a $k = 2$ stellar wind profile. Such a density profile could arise from a changing mass loss rate of the progenitor star as it reaches the end stages of its life. Standard afterglow models predict that the evolution of $\nu_{\textrm{m}}$ is independent of the circumburst environment, therefore we cannot assume that the slow evolution of $\nu_{\textrm{m}}$ is due to environmental effects. In other GRBs, \citep[e.g.][]{2019MNRAS.486.2721B} the unexpected evolution of $\nu_{\textrm{m}}$ is considered as a result of time-varying microphysical parameters or scintillation. In the case of GRB 221009A, we find no evidence for significant scintillation effects, and time-varying microphysical parameters would cause further changes in the evolution of $F_{\nu, \textrm{max}}$ and $\nu_{\textrm{sa}}$, which could potentially provide an alternative explanation, other than a steep $k$ value, for the observed behaviour. 

\subsubsection{Late-time evolution}

Our latest observations were made with ATCA at 475\,days post burst at 5.5 and 9\,GHz. Our model finds that at such late times, the forward shock is the brightest emission component during the decay phase of the light curve at these radio frequencies. Many late-time radio and X-ray light curves extending out to hundreds of days show achromatic behaviour referred to as a jet break \citep[e.g.][]{2010ApJ...725..625T,2021ApJ...911...14K}. As the jet decelerates, the beaming angle, dictating the fraction of jet that the observer can see, increases. Before the jet break, the light curve at a given frequency will decay at a shallower rate than the intrinsic evolution because a greater fraction of the jet is visible at every new time step. At the point where the opening angle is equal to the inverse of the bulk Lorentz factor, the jet break, the whole jet is within the beaming angle, so the light curve at all wavelengths will begin to decay at a steeper rate \citep[$t^{-3p/4}$ or $t^{-p}$, depending on whether lateral spreading is assumed or not,][]{1999ApJ...519L..17S, gao2013} which matches the intrinsic evolution of the shock. By observing the jet break, it is possible to measure the opening angle of the jet.

Jet breaks have been observed at many different times post-burst, from a fraction of a day to tens of days or even later. For most GRBs, the afterglow quickly fades below detection limits before a jet break can be observed. In some long-lasting afterglows, no jet break is observed at all for a very long time, the best example being GRB 130427A where no jet break was observed out to at least 1.9\,years post-burst \citep[][]{2016MNRAS.462.1111D}. Comparatively, we rule out the presence of a break in the light curve out to 1.3\,years based on our latest ATCA observations. We note that the presence of lateral structure, as indicated by the need for a third shock component which is discussed in Section \ref{sec:extra}, could disguise the jet break signature which is predicted for top hat jets \citep{1999ApJ...519L..17S, gao2013}.

Whilst the presence of the jet break is used to measure the jet opening angle, the measurement is also dependent on the jet's kinetic energy and the density of the circumburst environment. The fact that there has been no change in light curve behaviour out to over a year post-burst due to a jet break indicates that the kinetic energy of the jet could be higher than what is deemed `normal' for a regular GRB jet, the circumburst density is very low, or a it has a wide jet opening angle. As already suggested by \citet{oconnor2023}, GRB 221009A may belong to a class of hyper-energetic GRBs \citep{2008ApJ...683..924C, 2011ApJ...732...29C,2014A&A...567A..84M}, events whose kinetic energies are greater than $10^{51}$erg. Given the large isotropic-equivalent kinetic energies inferred from modelling so far, a large jet opening angle is unlikely as it would require the beaming-corrected kinetic energy to be physically challenging, approaching that of the isotopic equivalent kinetic energy. It has been suggested \citep[e.g.,][]{levan2023, oconnor2023} that a jet break occurred within the first day post-burst. Our observations and modelling provide no evidence that such a jet break occurred. 

There is also expected to be a change in the observed light curve behaviour as the jet leaves the stellar wind bubble produced by the progenitor star and enters the surrounding homogeneous interstellar medium. The stellar wind bubble is expected to be several tens of parsecs in size \citep{2005ApJ...630..892D, 2006MNRAS.367..186E}. For GRB 130427A, a stellar wind to homogeneous transition is ruled out to 1.9\,years post-burst.  In that case, it was estimated that the jet had travelled between 50 and 105\,parsecs, putting strong constraints on the presence/size of a termination shock, other nearby stars, etc. Our model for GRB 221009A disfavours any change in the structure of the circumburst environment out to 1.3\,years, or that the stellar wind bubble produced by the stellar progenitor exists in a very low pre-existing ISM density for the stellar wind to expand into. However, if the circumburst density profile is very steep, as our forward shock model suggests, it may be very difficult to observe such a transition. \citet{2011ApJ...732...29C} suggested that the hyper-energetic events can occur in lower metallicity environments where the progenitor star maintains a higher angular momentum for longer and therefore evacuates a larger cavity with its stellar wind, therefore delaying any change in temporal behaviour.

Studies of GRB progenitor systems predict termination shock radii to be less than 20\,parsec \citep{2006ApJ...647.1269F,2011A&A...526A..23S}. Using the radio source size growth rate from \citet{Giarratana2023}, we estimate the distance travelled by the jet for three different assumed opening angles. For opening angles of 2, 5 and 10$^{\circ}$, the jet should have propagated $\sim10$, 4 and 2\,parsecs, respectively. At the current epoch, our observations are still consistent with the sizes of termination shocks found in the literature \citep{2006ApJ...647.1269F}. Therefore, we can treat these values as lower limits on the termination shock size. Continued low-frequency radio observations will be vital in tracking the jet as it continues to expand into the surrounding medium.

\subsection{Extra shock}\label{sec:extra}

As explained in Section \ref{sec:model}, we ran two different iterations of the third shock component in our model to test different theoretical predictions (see Table \ref{tab:extra_shock} for a summary, and Figure \ref{fig:diff_models} for the results). The dash-dotted lines in Figures \ref{fig:lightcurve_model} and \ref{fig:sed_model} denote the contribution of this component. The most important aspect of the third spectral component is the delayed deceleration timescale over which the component comes into the observer's line of sight \citep{2005ApJ...626..966P}. We find that the deceleration time for the third component is 0.27$\pm$0.02\,days, the break time in our $F_{\nu, \textrm{max}}$ broken power law evolution. The delayed deceleration timescale is used to show that there is a possibility that the third component is either off-axis and therefore takes time to enter our line of sight, or that it is less relativistic than the main jet component and so needs longer to shock sufficient mass such that it undergoes significant deceleration.

To ensure that the data needs the $F_{\nu, \textrm{max}} \propto t^{3}$ rise, we also ran a separate model iteration which allows the rise index to vary (model 2 in Table \ref{tab:extra_shock}). In this iteration, we find a broad posterior distribution, i.e., not a Gaussian posterior, extending from $F_{\nu, \textrm{max}} \propto t^{1.6}$ to the edge of the prior which is $F_{\nu, \textrm{max}} \propto t^{3}$. Such a broad posterior could be indicative of some lateral structure in the outflow such that the whole shock front does not enter our line of sight at once \citep{2018ApJ...868L..11M,2020ApJ...896..166R}.

After the peak, for a decelerating shock, afterglow models predict $F_{\nu, \textrm{max}}$ to decay between $t^{-1.7}$ and $t^{-1.8}$, for $p = 3.1$ for a stellar wind and homogeneous medium, respectively \citep{granot2002}. Our observations find $F_{\nu, \textrm{max}} \propto t^{-0.71\pm0.03}$, significantly slower than the models predict. The break frequency $\nu_{\textrm{sa}}$ is expected to decay as  $t^{-1.1}$ and $t^{-1.3}$, for $k = 0$ and $2$, respectively, whereas we find $t^{-0.46^{+0.03}_{-0.04}}$. Therefore, we find that the evolution of $\nu_{\textrm{sa}}$ for this extra component is far slower than predicted by analytical blast wave models, contrary to the evolution in the forward shock case, which is too fast. 

We can also use the observed evolution to extract the density profile of the circumburst environment and $p$, independently of the spectral fit \citep{vanderhorst2014}. In this case, we take the observed $F_{\nu, \textrm{max}}$ and $\nu_{\textrm{sa}}$ behaviour as a function of time and solve for $p$ and $k$. However, solving for $p$ and $k$ does not provide physical solutions for either value, i.e., a negative value of $p$.

Given the clear disagreement between our modelling results using three components and expectations from analytical shock models, it is possible that this third additional component is not produced by a relativistic shock but a slower outflow component such a circumstellar interaction from the supernova. The peak luminosity of the extra shock is around $10^{30}$\,erg\,s\textsuperscript{-1}\,Hz\textsuperscript{-1} which is still an order of magnitude higher than the most luminous radio-detected supernovae \citep[e.g.][]{2019ApJ...872..201P,2021Sci...373.1125D}, and reaches such high luminosities within a day as opposed to 100-1000\,s of days later.

Therefore, we find that the origin of the additional spectral component is most likely a wider outflow or cocoon-like component, as opposed to circumstellar interaction from a supernova. Being slightly less relativistic than the jet, the cocoon will take less time to sweep up mass whose rest mass energy is equal to that of the outflow and therefore will experience delayed deceleration. It is also likely to be slightly off-axis compared to the forward and reverse shock-emitting jet. 

Cocoons have been invoked in previous GRB systems \citep[e.g.][]{2018Natur.554..207M,2019Natur.565..324I} where sufficiently high-quality data has been used to infer their presence. It is possible that cocoons are a more universal component of GRBs but our observations have been too sparse to find them.

\section{Conclusions}\label{sec:conc}

In this work, we have collated and presented the most detailed radio study of any long GRB to date. When combined with the published X-ray data, we find that the radio observations are best described with three synchrotron spectra, each evolving individually. A reverse shock component dominates the early-time low-frequency data below 20\,GHz. The higher-frequency radio emission and X-ray data can be ascribed to a forward shock. Due to the high temporal and spectral coverage, we are also able to constrain the evolution and properties of a third component which we attribute to a potential cocoon-like outflow. Whilst it is possible to match the different spectra with different shock components, we find that in all cases the evolution of the self-absorbed regions of the afterglow does not match up with the models currently in the literature. Also the peak frequency and peak flux show temporal behavior that is inconsistent with theoretical afterglow models. Given the high signal-to-noise of our latest observations, we aim to continue observing the afterglow of GRB 221009A for years to come to detect a potential jet break, track the jet into the non-relativistic regime, and constrain the size of the wind bubble in which this GRB resides.

\section*{Acknowledgements}

The authors would like the thank the anonymous referee for their helpful comments. L.R. thanks Eric Burns, Courey Elliott, Geoffrey Ryan and Ashley Villar for useful conversations in the writing of this paper. K.G. acknowledges support through the Australian Research Council Discovery Project DP200102243. M.J.M.~acknowledges the support of the National Science Centre, Poland through the SONATA BIS grant 2018/30/E/ST9/00208 and the Polish National Agency for Academic Exchange Bekker grant BPN/BEK/2022/1/00110. AR acknowledges funding from the NOW Aspasia grant (number: 015.016.033). RLCS acknowledges support from a Leverhulme Trust Research Project Grant.

Parts of this research were conducted by the Australian Research Council Centre of Excellence for Gravitational Wave Discovery (OzGrav), project numbers CE170100004 and CE230100016. We thank the Mullard Radio Astronomy Observatory staff for scheduling and carrying out the AMI-LA observations. The AMI telescope is supported by the European Research Council under grant ERC2012-StG-307215 LODESTONE, the UK Science and Technology
Facilities Council, and the University of Cambridge. The Allen Telescope Array refurbishment program and its ongoing operations are being substantially funded through the Franklin Antonio Bequest. Additional contributions from Frank Levinson, Greg Papadopoulos, the Breakthrough Listen Initiative and other private donors have been instrumental in the renewal of the ATA. Breakthrough Listen is managed by the Breakthrough Initiatives, sponsored by the Breakthrough Prize Foundation. The Paul G. Allen Family Foundation provided major support for the design and construction of the ATA, alongside contributions from Nathan Myhrvold, Xilinx Corporation, Sun Microsystems, and other private donors. The ATA has also been supported by contributions from the US Naval Observatory and the US National Science Foundation. The Australia Telescope Compact Array is part of the Australia Telescope National Facility\footnote{https://ror.org/05qajvd42} which is funded by the Australian Government for operation as a National Facility managed by CSIRO. We acknowledge the Gomeroi people as the Traditional Owners of the Observatory site. This scientific work uses data obtained from Inyarrimanha Ilgari Bundara / the Murchison Radio-astronomy Observatory. We acknowledge the Wajarri Yamaji People as the Traditional Owners and native title holders of the Observatory site. CSIRO’s ASKAP radio telescope is part of the Australia Telescope National Facility\footnotemark[\value{footnote}]. Operation of ASKAP is funded by the Australian Government with support from the National Collaborative Research Infrastructure Strategy. ASKAP uses the resources of the Pawsey Supercomputing Research Centre. Establishment of ASKAP, Inyarrimanha Ilgari Bundara, the CSIRO Murchison Radio-astronomy Observatory and the Pawsey Supercomputing Research Centre are initiatives of the Australian Government, with support from the Government of Western Australia and the Science and Industry Endowment Fund. e-MERLIN is a National Facility operated by the University of Manchester at Jodrell Bank Observatory on behalf of STFC. This paper is based (in part) on data obtained with the International LOFAR Telescope (ILT) under project code DDT20\_003. LOFAR \citep{2013A&A...556A...2V} is the Low Frequency Array designed and constructed by ASTRON. It has observing, data processing, and data storage facilities in several countries, that are owned by various parties (each with their own funding sources), and that are collectively operated by the ILT foundation under a joint scientific policy. The ILT resources have benefited from the following recent major funding sources: CNRS-INSU, Observatoire de Paris and Université d'Orléans, France; BMBF, MIWF-NRW, MPG, Germany; Science Foundation Ireland (SFI), Department of Business, Enterprise and Innovation (DBEI), Ireland; NWO, The Netherlands; The Science and Technology Facilities Council, UK; Ministry of Science and Higher Education, Poland. Based on observations carried out under project numbers S22BC, S22BF, W22BU and S22BE with the IRAM NOEMA Interferometer. IRAM is supported by INSU/CNRS (France), MPG (Germany) and IGN (Spain). We thank the staff of the GMRT that made these observations possible. GMRT is run by the National Centre for Radio Astrophysics of the Tata Institute of Fundamental Research. This work made use of data supplied by the UK Swift Science Data Centre at the University of Leicester. The National Radio Astronomy Observatory is a facility of the National Science Foundation operated under cooperative agreement by Associated Universities, Inc.

\section*{Data Availability}

All the new data presented in this paper is given the supplementary material.



\bibliographystyle{mnras}
\bibliography{example} 

\begin{thebibliography}{}
\makeatletter
\relax
\def\mn@urlcharsother{\let\do\@makeother \do\$\do\&\do\#\do\^\do\_\do\%\do\~}
\def\mn@doi{\begingroup\mn@urlcharsother \@ifnextchar [ {\mn@doi@} {\mn@doi@[]}}
\def\mn@doi@[#1]#2{\def\@tempa{#1}\ifx\@tempa\@empty \href {http://dx.doi.org/#2} {doi:#2}\else \href {http://dx.doi.org/#2} {#1}\fi \endgroup}
\def\mn@eprint#1#2{\mn@eprint@#1:#2::\@nil}
\def\mn@eprint@arXiv#1{\href {http://arxiv.org/abs/#1} {{\tt arXiv:#1}}}
\def\mn@eprint@dblp#1{\href {http://dblp.uni-trier.de/rec/bibtex/#1.xml} {dblp:#1}}
\def\mn@eprint@#1:#2:#3:#4\@nil{\def\@tempa {#1}\def\@tempb {#2}\def\@tempc {#3}\ifx \@tempc \@empty \let \@tempc \@tempb \let \@tempb \@tempa \fi \ifx \@tempb \@empty \def\@tempb {arXiv}\fi \@ifundefined {mn@eprint@\@tempb}{\@tempb:\@tempc}{\expandafter \expandafter \csname mn@eprint@\@tempb\endcsname \expandafter{\@tempc}}}

\bibitem[\protect\citeauthoryear{{Abe} et~al.,}{{Abe} et~al.}{2024}]{2024MNRAS.527.5856A}
{Abe} H.,  et~al., 2024, \mn@doi [\mnras] {10.1093/mnras/stad2958}, \href {https://ui.adsabs.harvard.edu/abs/2024MNRAS.527.5856A} {527, 5856}

\bibitem[\protect\citeauthoryear{{Achterberg}, {Gallant}, {Kirk}  \& {Guthmann}}{{Achterberg} et~al.}{2001}]{2001MNRAS.328..393A}
{Achterberg} A.,  {Gallant} Y.~A.,  {Kirk} J.~G.,   {Guthmann} A.~W.,  2001, \mn@doi [\mnras] {10.1046/j.1365-8711.2001.04851.x}, \href {https://ui.adsabs.harvard.edu/abs/2001MNRAS.328..393A} {328, 393}

\bibitem[\protect\citeauthoryear{{Ackermann} et~al.,}{{Ackermann} et~al.}{2014}]{2014Sci...343...42A}
{Ackermann} M.,  et~al., 2014, \mn@doi [Science] {10.1126/science.1242353}, \href {https://ui.adsabs.harvard.edu/abs/2014Sci...343...42A} {343, 42}

\bibitem[\protect\citeauthoryear{{Aharonian} et~al.,}{{Aharonian} et~al.}{2023}]{2023ApJ...946L..27A}
{Aharonian} F.,  et~al., 2023, \mn@doi [\apjl] {10.3847/2041-8213/acc405}, \href {https://ui.adsabs.harvard.edu/abs/2023ApJ...946L..27A} {946, L27}

\bibitem[\protect\citeauthoryear{{An} et~al.,}{{An} et~al.}{2023}]{2023arXiv230301203A}
{An} Z.-H.,  et~al., 2023, \mn@doi [arXiv e-prints] {10.48550/arXiv.2303.01203}, \href {https://ui.adsabs.harvard.edu/abs/2023arXiv230301203A} {p. arXiv:2303.01203}

\bibitem[\protect\citeauthoryear{{Anderson} et~al.,}{{Anderson} et~al.}{2014}]{2014MNRAS.440.2059A}
{Anderson} G.~E.,  et~al., 2014, \mn@doi [\mnras] {10.1093/mnras/stu478}, \href {https://ui.adsabs.harvard.edu/abs/2014MNRAS.440.2059A} {440, 2059}

\bibitem[\protect\citeauthoryear{{Blanchard} et~al.,}{{Blanchard} et~al.}{2024}]{2024NatAs.tmp...65B}
{Blanchard} P.~K.,  et~al., 2024, \mn@doi [Nature Astronomy] {10.1038/s41550-024-02237-4}, \href {https://ui.adsabs.harvard.edu/abs/2024NatAs.tmp...65B} {}

\bibitem[\protect\citeauthoryear{{Bright J. S. \& Rhodes L.} et~al.,}{{Bright J. S. \& Rhodes L.} et~al.}{2023}]{bright2023}
{Bright J. S. \& Rhodes L.} et~al., 2023, \mn@doi [Nature Astronomy] {10.1038/s41550-023-01997-9}, 7, 986

\bibitem[\protect\citeauthoryear{{Bright} et~al.,}{{Bright} et~al.}{2019}]{2019MNRAS.486.2721B}
{Bright} J.~S.,  et~al., 2019, \mn@doi [\mnras] {10.1093/mnras/stz1004}, \href {https://ui.adsabs.harvard.edu/abs/2019MNRAS.486.2721B} {486, 2721}

\bibitem[\protect\citeauthoryear{{Burns} et~al.,}{{Burns} et~al.}{2023}]{burns2023}
{Burns} E.,  et~al., 2023, \mn@doi [\apjl] {10.3847/2041-8213/acc39c}, 946, L31

\bibitem[\protect\citeauthoryear{{Cenko} et~al.,}{{Cenko} et~al.}{2011}]{2011ApJ...732...29C}
{Cenko} S.~B.,  et~al., 2011, \mn@doi [\apj] {10.1088/0004-637X/732/1/29}, \href {https://ui.adsabs.harvard.edu/abs/2011ApJ...732...29C} {732, 29}

\bibitem[\protect\citeauthoryear{{Chandra} et~al.,}{{Chandra} et~al.}{2008}]{2008ApJ...683..924C}
{Chandra} P.,  et~al., 2008, \mn@doi [\apj] {10.1086/589807}, \href {https://ui.adsabs.harvard.edu/abs/2008ApJ...683..924C} {683, 924}

\bibitem[\protect\citeauthoryear{{Chevalier} \& {Li}}{{Chevalier} \& {Li}}{1999}]{Chevalier1999}
{Chevalier} R.~A.,  {Li} Z.-Y.,  1999, \mn@doi [\apjl] {10.1086/312147}, \href {https://ui.adsabs.harvard.edu/abs/1999ApJ...520L..29C} {520, L29}

\bibitem[\protect\citeauthoryear{{Clark}}{{Clark}}{1980}]{Clark1980}
{Clark} B.~G.,  1980, \aap, \href {https://ui.adsabs.harvard.edu/abs/1980A&A....89..377C} {89, 377}

\bibitem[\protect\citeauthoryear{{De Pasquale} et~al.,}{{De Pasquale} et~al.}{2016}]{2016MNRAS.462.1111D}
{De Pasquale} M.,  et~al., 2016, \mn@doi [\mnras] {10.1093/mnras/stw1704}, \href {https://ui.adsabs.harvard.edu/abs/2016MNRAS.462.1111D} {462, 1111}

\bibitem[\protect\citeauthoryear{{Dong} et~al.,}{{Dong} et~al.}{2021}]{2021Sci...373.1125D}
{Dong} D.~Z.,  et~al., 2021, \mn@doi [Science] {10.1126/science.abg6037}, \href {https://ui.adsabs.harvard.edu/abs/2021Sci...373.1125D} {373, 1125}

\bibitem[\protect\citeauthoryear{{Dwarkadas}}{{Dwarkadas}}{2005}]{2005ApJ...630..892D}
{Dwarkadas} V.~V.,  2005, \mn@doi [\apj] {10.1086/432109}, \href {https://ui.adsabs.harvard.edu/abs/2005ApJ...630..892D} {630, 892}

\bibitem[\protect\citeauthoryear{{Eichler} \& {Waxman}}{{Eichler} \& {Waxman}}{2005}]{eichlergranot2005}
{Eichler} D.,  {Waxman} E.,  2005, \mn@doi [\apj] {10.1086/430596}, \href {https://ui.adsabs.harvard.edu/abs/2005ApJ...627..861E} {627, 861}

\bibitem[\protect\citeauthoryear{{Eldridge}, {Genet}, {Daigne}  \& {Mochkovitch}}{{Eldridge} et~al.}{2006}]{2006MNRAS.367..186E}
{Eldridge} J.~J.,  {Genet} F.,  {Daigne} F.,   {Mochkovitch} R.,  2006, \mn@doi [\mnras] {10.1111/j.1365-2966.2005.09938.x}, \href {https://ui.adsabs.harvard.edu/abs/2006MNRAS.367..186E} {367, 186}

\bibitem[\protect\citeauthoryear{{Foreman-Mackey}, {Hogg}, {Lang}  \& {Goodman}}{{Foreman-Mackey} et~al.}{2013}]{Foreman-Mackey2013}
{Foreman-Mackey} D.,  {Hogg} D.~W.,  {Lang} D.,   {Goodman} J.,  2013, \mn@doi [\pasp] {10.1086/670067}, \href {https://ui.adsabs.harvard.edu/abs/2013PASP..125..306F} {125, 306}

\bibitem[\protect\citeauthoryear{{Frail}, {Waxman}  \& {Kulkarni}}{{Frail} et~al.}{2000}]{2000ApJ...537..191F}
{Frail} D.~A.,  {Waxman} E.,   {Kulkarni} S.~R.,  2000, \mn@doi [\apj] {10.1086/309024}, \href {https://ui.adsabs.harvard.edu/abs/2000ApJ...537..191F} {537, 191}

\bibitem[\protect\citeauthoryear{{Frederiks} et~al.,}{{Frederiks} et~al.}{2023}]{2023ApJ...949L...7F}
{Frederiks} D.,  et~al., 2023, \mn@doi [\apjl] {10.3847/2041-8213/acd1eb}, \href {https://ui.adsabs.harvard.edu/abs/2023ApJ...949L...7F} {949, L7}

\bibitem[\protect\citeauthoryear{{Fryer}, {Rockefeller}  \& {Young}}{{Fryer} et~al.}{2006}]{2006ApJ...647.1269F}
{Fryer} C.~L.,  {Rockefeller} G.,   {Young} P.~A.,  2006, \mn@doi [\apj] {10.1086/505590}, \href {https://ui.adsabs.harvard.edu/abs/2006ApJ...647.1269F} {647, 1269}

\bibitem[\protect\citeauthoryear{{Fulton} et~al.,}{{Fulton} et~al.}{2023}]{Fulton2023}
{Fulton} M.~D.,  et~al., 2023, \mn@doi [\apjl] {10.3847/2041-8213/acc101}, \href {https://ui.adsabs.harvard.edu/abs/2023ApJ...946L..22F} {946, L22}

\bibitem[\protect\citeauthoryear{{Gao}, {Lei}, {Zou}, {Wu}  \& {Zhang}}{{Gao} et~al.}{2013}]{gao2013}
{Gao} H.,  {Lei} W.-H.,  {Zou} Y.-C.,  {Wu} X.-F.,   {Zhang} B.,  2013, \mn@doi [\nar] {10.1016/j.newar.2013.10.001}, 57, 141

\bibitem[\protect\citeauthoryear{{Giarratana} et~al.,}{{Giarratana} et~al.}{2023}]{Giarratana2023}
{Giarratana} S.,  et~al., 2023, \mn@doi [arXiv e-prints] {10.48550/arXiv.2311.05527}, \href {https://ui.adsabs.harvard.edu/abs/2023arXiv231105527G} {p. arXiv:2311.05527}

\bibitem[\protect\citeauthoryear{{Gill} \& {Granot}}{{Gill} \& {Granot}}{2023}]{2023MNRAS.524L..78G}
{Gill} R.,  {Granot} J.,  2023, \mn@doi [\mnras] {10.1093/mnrasl/slad075}, \href {https://ui.adsabs.harvard.edu/abs/2023MNRAS.524L..78G} {524, L78}

\bibitem[\protect\citeauthoryear{{Granot} \& {Sari}}{{Granot} \& {Sari}}{2002}]{granot2002}
{Granot} J.,  {Sari} R.,  2002, \mn@doi [\apj] {10.1086/338966}, \href {https://ui.adsabs.harvard.edu/abs/2002ApJ...568..820G} {568, 820}

\bibitem[\protect\citeauthoryear{{Granot} \& {van der Horst}}{{Granot} \& {van der Horst}}{2014}]{2014PASA...31....8G}
{Granot} J.,  {van der Horst} A.~J.,  2014, \mn@doi [\pasa] {10.1017/pasa.2013.44}, \href {https://ui.adsabs.harvard.edu/abs/2014PASA...31....8G} {31, e008}

\bibitem[\protect\citeauthoryear{{Greiner} et~al.,}{{Greiner} et~al.}{2009}]{2009A&A...498...89G}
{Greiner} J.,  et~al., 2009, \mn@doi [\aap] {10.1051/0004-6361/200811571}, \href {https://ui.adsabs.harvard.edu/abs/2009A&A...498...89G} {498, 89}

\bibitem[\protect\citeauthoryear{{Guzman} et~al.,}{{Guzman} et~al.}{2019}]{guzman2019}
{Guzman} J.,  et~al., 2019, {ASKAPsoft: ASKAP science data processor software}, Astrophysics Source Code Library, record ascl:1912.003 (\mn@eprint {ascl} {1912.003})

\bibitem[\protect\citeauthoryear{{H.~E.~S.~S. Collaboration} et~al.,}{{H.~E.~S.~S. Collaboration} et~al.}{2021}]{2021Sci...372.1081H}
{H.~E.~S.~S. Collaboration} et~al., 2021, \mn@doi [Science] {10.1126/science.abe8560}, \href {https://ui.adsabs.harvard.edu/abs/2021Sci...372.1081H} {372, 1081}

\bibitem[\protect\citeauthoryear{{Hale} et~al.,}{{Hale} et~al.}{2021}]{hale2021}
{Hale} C.~L.,  et~al., 2021, \mn@doi [\pasa] {10.1017/pasa.2021.47}, \href {https://ui.adsabs.harvard.edu/abs/2021PASA...38...58H} {38, e058}

\bibitem[\protect\citeauthoryear{{Hayes} \& {Gallagher}}{{Hayes} \& {Gallagher}}{2022}]{2022RNAAS...6..222H}
{Hayes} L.~A.,  {Gallagher} P.~T.,  2022, \mn@doi [Research Notes of the American Astronomical Society] {10.3847/2515-5172/ac9d2f}, \href {https://ui.adsabs.harvard.edu/abs/2022RNAAS...6..222H} {6, 222}

\bibitem[\protect\citeauthoryear{{Hickish} et~al.,}{{Hickish} et~al.}{2018}]{2018MNRAS.475.5677H}
{Hickish} J.,  et~al., 2018, \mn@doi [\mnras] {10.1093/mnras/sty074}, \href {https://ui.adsabs.harvard.edu/abs/2018MNRAS.475.5677H} {475, 5677}

\bibitem[\protect\citeauthoryear{{H{\"o}gbom}}{{H{\"o}gbom}}{1974}]{Hogbom1974}
{H{\"o}gbom} J.~A.,  1974, \aaps, \href {https://ui.adsabs.harvard.edu/abs/1974A&AS...15..417H} {15, 417}

\bibitem[\protect\citeauthoryear{{Hotan} et~al.,}{{Hotan} et~al.}{2021}]{hotan2021}
{Hotan} A.~W.,  et~al., 2021, \mn@doi [\pasa] {10.1017/pasa.2021.1}, \href {https://ui.adsabs.harvard.edu/abs/2021PASA...38....9H} {38, e009}

\bibitem[\protect\citeauthoryear{{Izzo} et~al.,}{{Izzo} et~al.}{2019}]{2019Natur.565..324I}
{Izzo} L.,  et~al., 2019, \mn@doi [\nat] {10.1038/s41586-018-0826-3}, \href {https://ui.adsabs.harvard.edu/abs/2019Natur.565..324I} {565, 324}

\bibitem[\protect\citeauthoryear{{Johnston} et~al.,}{{Johnston} et~al.}{2007}]{Johnston2007}
{Johnston} S.,  et~al., 2007, \mn@doi [\pasa] {10.1071/AS07033}, \href {https://ui.adsabs.harvard.edu/abs/2007PASA...24..174J} {24, 174}

\bibitem[\protect\citeauthoryear{{Kangas} \& {Fruchter}}{{Kangas} \& {Fruchter}}{2021}]{2021ApJ...911...14K}
{Kangas} T.,  {Fruchter} A.~S.,  2021, \mn@doi [\apj] {10.3847/1538-4357/abe76b}, \href {https://ui.adsabs.harvard.edu/abs/2021ApJ...911...14K} {911, 14}

\bibitem[\protect\citeauthoryear{{Kann} et~al.,}{{Kann} et~al.}{2023}]{2023ApJ...948L..12K}
{Kann} D.~A.,  et~al., 2023, \mn@doi [\apjl] {10.3847/2041-8213/acc8d0}, \href {https://ui.adsabs.harvard.edu/abs/2023ApJ...948L..12K} {948, L12}

\bibitem[\protect\citeauthoryear{{Kirk}, {Guthmann}, {Gallant}  \& {Achterberg}}{{Kirk} et~al.}{2000}]{2000ApJ...542..235K}
{Kirk} J.~G.,  {Guthmann} A.~W.,  {Gallant} Y.~A.,   {Achterberg} A.,  2000, \mn@doi [\apj] {10.1086/309533}, \href {https://ui.adsabs.harvard.edu/abs/2000ApJ...542..235K} {542, 235}

\bibitem[\protect\citeauthoryear{{Kobayashi} \& {Sari}}{{Kobayashi} \& {Sari}}{2000}]{2000ApJ...542..819K}
{Kobayashi} S.,  {Sari} R.,  2000, \mn@doi [\apj] {10.1086/317021}, \href {https://ui.adsabs.harvard.edu/abs/2000ApJ...542..819K} {542, 819}

\bibitem[\protect\citeauthoryear{{LHAASO Collaboration} et~al.,}{{LHAASO Collaboration} et~al.}{2023}]{lhaaso2023}
{LHAASO Collaboration} et~al., 2023, \mn@doi [Science] {10.1126/science.adg9328}, 380, 1390

\bibitem[\protect\citeauthoryear{{Lamb}, {Kann}, {Fern{\'a}ndez}, {Mandel}, {Levan}  \& {Tanvir}}{{Lamb} et~al.}{2021}]{2021MNRAS.506.4163L}
{Lamb} G.~P.,  {Kann} D.~A.,  {Fern{\'a}ndez} J.~J.,  {Mandel} I.,  {Levan} A.~J.,   {Tanvir} N.~R.,  2021, \mn@doi [\mnras] {10.1093/mnras/stab2071}, \href {https://ui.adsabs.harvard.edu/abs/2021MNRAS.506.4163L} {506, 4163}

\bibitem[\protect\citeauthoryear{{Laskar} et~al.,}{{Laskar} et~al.}{2019}]{2019ApJ...878L..26L}
{Laskar} T.,  et~al., 2019, \mn@doi [\apjl] {10.3847/2041-8213/ab2247}, \href {https://ui.adsabs.harvard.edu/abs/2019ApJ...878L..26L} {878, L26}

\bibitem[\protect\citeauthoryear{{Laskar} et~al.,}{{Laskar} et~al.}{2023}]{laskar2023}
{Laskar} T.,  et~al., 2023, \mn@doi [\apjl] {10.3847/2041-8213/acbfad}, \href {https://ui.adsabs.harvard.edu/abs/2023ApJ...946L..23L} {946, L23}

\bibitem[\protect\citeauthoryear{{Lesage} et~al.,}{{Lesage} et~al.}{2023}]{lesage2023}
{Lesage} S.,  et~al., 2023, \mn@doi [\apjl] {10.3847/2041-8213/ace5b4}, 952, L42

\bibitem[\protect\citeauthoryear{{Levan} et~al.,}{{Levan} et~al.}{2014}]{2014ApJ...792..115L}
{Levan} A.~J.,  et~al., 2014, \mn@doi [\apj] {10.1088/0004-637X/792/2/115}, \href {https://ui.adsabs.harvard.edu/abs/2014ApJ...792..115L} {792, 115}

\bibitem[\protect\citeauthoryear{{Levan} et~al.,}{{Levan} et~al.}{2023}]{levan2023}
{Levan} A.~J.,  et~al., 2023, \mn@doi [\apjl] {10.3847/2041-8213/acc2c1}, \href {https://ui.adsabs.harvard.edu/abs/2023ApJ...946L..28L} {946, L28}

\bibitem[\protect\citeauthoryear{{MAGIC Collaboration} et~al.,}{{MAGIC Collaboration} et~al.}{2019a}]{2019Natur.575..455M}
{MAGIC Collaboration} et~al., 2019a, \mn@doi [\nat] {10.1038/s41586-019-1750-x}, \href {https://ui.adsabs.harvard.edu/abs/2019Natur.575..455M} {575, 455}

\bibitem[\protect\citeauthoryear{{MAGIC Collaboration} et~al.,}{{MAGIC Collaboration} et~al.}{2019b}]{2019Natur.575..459M}
{MAGIC Collaboration} et~al., 2019b, \mn@doi [\nat] {10.1038/s41586-019-1754-6}, \href {https://ui.adsabs.harvard.edu/abs/2019Natur.575..459M} {575, 459}

\bibitem[\protect\citeauthoryear{{Malesani} et~al.,}{{Malesani} et~al.}{2023}]{2023arXiv230207891M}
{Malesani} D.~B.,  et~al., 2023, \mn@doi [arXiv e-prints] {10.48550/arXiv.2302.07891}, \href {https://ui.adsabs.harvard.edu/abs/2023arXiv230207891M} {p. arXiv:2302.07891}

\bibitem[\protect\citeauthoryear{{Martin-Carrillo} et~al.,}{{Martin-Carrillo} et~al.}{2014}]{2014A&A...567A..84M}
{Martin-Carrillo} A.,  et~al., 2014, \mn@doi [\aap] {10.1051/0004-6361/201220872}, \href {https://ui.adsabs.harvard.edu/abs/2014A&A...567A..84M} {567, A84}

\bibitem[\protect\citeauthoryear{{McMullin}, {Waters}, {Schiebel}, {Young}  \& {Golap}}{{McMullin} et~al.}{2007}]{McMullin2007}
{McMullin} J.~P.,  {Waters} B.,  {Schiebel} D.,  {Young} W.,   {Golap} K.,  2007, in {Shaw} R.~A.,  {Hill} F.,   {Bell} D.~J.,  eds,  Astronomical Society of the Pacific Conference Series Vol. 376, Astronomical Data Analysis Software and Systems XVI. p.~127

\bibitem[\protect\citeauthoryear{{Meszaros} \& {Rees}}{{Meszaros} \& {Rees}}{1993}]{1993ApJ...405..278M}
{Meszaros} P.,  {Rees} M.~J.,  1993, \mn@doi [\apj] {10.1086/172360}, \href {https://ui.adsabs.harvard.edu/abs/1993ApJ...405..278M} {405, 278}

\bibitem[\protect\citeauthoryear{{M{\'e}sz{\'a}ros} \& {Rees}}{{M{\'e}sz{\'a}ros} \& {Rees}}{1999}]{1999MNRAS.306L..39M}
{M{\'e}sz{\'a}ros} P.,  {Rees} M.~J.,  1999, \mn@doi [\mnras] {10.1046/j.1365-8711.1999.02800.x}, \href {https://ui.adsabs.harvard.edu/abs/1999MNRAS.306L..39M} {306, L39}

\bibitem[\protect\citeauthoryear{{Misra} et~al.,}{{Misra} et~al.}{2021}]{2021MNRAS.504.5685M}
{Misra} K.,  et~al., 2021, \mn@doi [\mnras] {10.1093/mnras/stab1050}, \href {https://ui.adsabs.harvard.edu/abs/2021MNRAS.504.5685M} {504, 5685}

\bibitem[\protect\citeauthoryear{{Moldon}}{{Moldon}}{2021}]{2021ascl.soft09006M}
{Moldon} J.,  2021, {eMCP: e-MERLIN CASA pipeline}, Astrophysics Source Code Library, record ascl:2109.006 (\mn@eprint {ascl} {2109.006})

\bibitem[\protect\citeauthoryear{{Mooley} et~al.,}{{Mooley} et~al.}{2018a}]{2018Natur.554..207M}
{Mooley} K.~P.,  et~al., 2018a, \mn@doi [\nat] {10.1038/nature25452}, \href {https://ui.adsabs.harvard.edu/abs/2018Natur.554..207M} {554, 207}

\bibitem[\protect\citeauthoryear{{Mooley} et~al.,}{{Mooley} et~al.}{2018b}]{2018ApJ...868L..11M}
{Mooley} K.~P.,  et~al., 2018b, \mn@doi [\apjl] {10.3847/2041-8213/aaeda7}, \href {https://ui.adsabs.harvard.edu/abs/2018ApJ...868L..11M} {868, L11}

\bibitem[\protect\citeauthoryear{{Nakar} \& {Piran}}{{Nakar} \& {Piran}}{2017}]{2017ApJ...834...28N}
{Nakar} E.,  {Piran} T.,  2017, \mn@doi [\apj] {10.3847/1538-4357/834/1/28}, \href {https://ui.adsabs.harvard.edu/abs/2017ApJ...834...28N} {834, 28}

\bibitem[\protect\citeauthoryear{{Nayana}, {Chandra}, {Krishna}  \& {Anupama}}{{Nayana} et~al.}{2022}]{2022ApJ...934..186N}
{Nayana} A.~J.,  {Chandra} P.,  {Krishna} A.,   {Anupama} G.~C.,  2022, \mn@doi [\apj] {10.3847/1538-4357/ac7c1e}, \href {https://ui.adsabs.harvard.edu/abs/2022ApJ...934..186N} {934, 186}

\bibitem[\protect\citeauthoryear{{O'Connor} et~al.,}{{O'Connor} et~al.}{2023}]{oconnor2023}
{O'Connor} B.,  et~al., 2023, \mn@doi [Science Advances] {10.1126/sciadv.adi1405}, \href {https://ui.adsabs.harvard.edu/abs/2023SciA....9I1405O} {9, eadi1405}

\bibitem[\protect\citeauthoryear{{Offringa}}{{Offringa}}{2010}]{aoflagger}
{Offringa} A.~R.,  2010, {AOFlagger: RFI Software}, Astrophysics Source Code Library, record ascl:1010.017 (\mn@eprint {ascl} {1010.017})

\bibitem[\protect\citeauthoryear{{Palliyaguru} et~al.,}{{Palliyaguru} et~al.}{2019}]{2019ApJ...872..201P}
{Palliyaguru} N.~T.,  et~al., 2019, \mn@doi [\apj] {10.3847/1538-4357/aaf64d}, \href {https://ui.adsabs.harvard.edu/abs/2019ApJ...872..201P} {872, 201}

\bibitem[\protect\citeauthoryear{{Peng}, {K{\"o}nigl}  \& {Granot}}{{Peng} et~al.}{2005}]{2005ApJ...626..966P}
{Peng} F.,  {K{\"o}nigl} A.,   {Granot} J.,  2005, \mn@doi [\apj] {10.1086/430045}, \href {https://ui.adsabs.harvard.edu/abs/2005ApJ...626..966P} {626, 966}

\bibitem[\protect\citeauthoryear{{Perley} et~al.,}{{Perley} et~al.}{2014}]{perley2014}
{Perley} D.~A.,  et~al., 2014, \mn@doi [\apj] {10.1088/0004-637X/781/1/37}, \href {https://ui.adsabs.harvard.edu/abs/2014ApJ...781...37P} {781, 37}

\bibitem[\protect\citeauthoryear{{Perrott} et~al.,}{{Perrott} et~al.}{2013}]{2013MNRAS.429.3330P}
{Perrott} Y.~C.,  et~al., 2013, \mn@doi [\mnras] {10.1093/mnras/sts589}, \href {https://ui.adsabs.harvard.edu/abs/2013MNRAS.429.3330P} {429, 3330}

\bibitem[\protect\citeauthoryear{{Ramirez-Ruiz}, {Celotti}  \& {Rees}}{{Ramirez-Ruiz} et~al.}{2002}]{ramirezruiz2002}
{Ramirez-Ruiz} E.,  {Celotti} A.,   {Rees} M.~J.,  2002, \mn@doi [\mnras] {10.1046/j.1365-8711.2002.05995.x}, \href {https://ui.adsabs.harvard.edu/abs/2002MNRAS.337.1349R} {337, 1349}

\bibitem[\protect\citeauthoryear{{Resmi} et~al.,}{{Resmi} et~al.}{2005}]{2005A&A...440..477R}
{Resmi} L.,  et~al., 2005, \mn@doi [\aap] {10.1051/0004-6361:20041642}, \href {https://ui.adsabs.harvard.edu/abs/2005A&A...440..477R} {440, 477}

\bibitem[\protect\citeauthoryear{{Rhodes} et~al.,}{{Rhodes} et~al.}{2020}]{2020MNRAS.496.3326R}
{Rhodes} L.,  et~al., 2020, \mn@doi [\mnras] {10.1093/mnras/staa1715}, \href {https://ui.adsabs.harvard.edu/abs/2020MNRAS.496.3326R} {496, 3326}

\bibitem[\protect\citeauthoryear{{Rhodes}, {van der Horst}, {Fender}, {Aguilera-Dena}, {Bright}, {Vergani}  \& {Williams}}{{Rhodes} et~al.}{2022}]{2022MNRAS.513.1895R}
{Rhodes} L.,  {van der Horst} A.~J.,  {Fender} R.,  {Aguilera-Dena} D.~R.,  {Bright} J.~S.,  {Vergani} S.,   {Williams} D.~R.~A.,  2022, \mn@doi [\mnras] {10.1093/mnras/stac1057}, \href {https://ui.adsabs.harvard.edu/abs/2022MNRAS.513.1895R} {513, 1895}

\bibitem[\protect\citeauthoryear{{Rossi} \& {Rees}}{{Rossi} \& {Rees}}{2003}]{2003MNRAS.339..881R}
{Rossi} E.,  {Rees} M.~J.,  2003, \mn@doi [\mnras] {10.1046/j.1365-8711.2003.06242.x}, \href {https://ui.adsabs.harvard.edu/abs/2003MNRAS.339..881R} {339, 881}

\bibitem[\protect\citeauthoryear{{Rossi}, {Lazzati}  \& {Rees}}{{Rossi} et~al.}{2002}]{rossi2002}
{Rossi} E.,  {Lazzati} D.,   {Rees} M.~J.,  2002, \mn@doi [\mnras] {10.1046/j.1365-8711.2002.05363.x}, \href {https://ui.adsabs.harvard.edu/abs/2002MNRAS.332..945R} {332, 945}

\bibitem[\protect\citeauthoryear{{Ryan}, {van Eerten}, {Piro}  \& {Troja}}{{Ryan} et~al.}{2020}]{2020ApJ...896..166R}
{Ryan} G.,  {van Eerten} H.,  {Piro} L.,   {Troja} E.,  2020, \mn@doi [\apj] {10.3847/1538-4357/ab93cf}, \href {https://ui.adsabs.harvard.edu/abs/2020ApJ...896..166R} {896, 166}

\bibitem[\protect\citeauthoryear{{Salafia} \& {Ghirlanda}}{{Salafia} \& {Ghirlanda}}{2022}]{2022Galax..10...93S}
{Salafia} O.~S.,  {Ghirlanda} G.,  2022, \mn@doi [Galaxies] {10.3390/galaxies10050093}, \href {https://ui.adsabs.harvard.edu/abs/2022Galax..10...93S} {10, 93}

\bibitem[\protect\citeauthoryear{{Salafia} et~al.,}{{Salafia} et~al.}{2022}]{2022ApJ...931L..19S}
{Salafia} O.~S.,  et~al., 2022, \mn@doi [\apjl] {10.3847/2041-8213/ac6c28}, \href {https://ui.adsabs.harvard.edu/abs/2022ApJ...931L..19S} {931, L19}

\bibitem[\protect\citeauthoryear{{Sari} \& {Piran}}{{Sari} \& {Piran}}{1995}]{1995ApJ...455L.143S}
{Sari} R.,  {Piran} T.,  1995, \mn@doi [\apjl] {10.1086/309835}, \href {https://ui.adsabs.harvard.edu/abs/1995ApJ...455L.143S} {455, L143}

\bibitem[\protect\citeauthoryear{{Sari}, {Piran}  \& {Narayan}}{{Sari} et~al.}{1998}]{sari1998}
{Sari} R.,  {Piran} T.,   {Narayan} R.,  1998, \mn@doi [\apjl] {10.1086/311269}, \href {https://ui.adsabs.harvard.edu/abs/1998ApJ...497L..17S} {497, L17}

\bibitem[\protect\citeauthoryear{{Sari}, {Piran}  \& {Halpern}}{{Sari} et~al.}{1999}]{1999ApJ...519L..17S}
{Sari} R.,  {Piran} T.,   {Halpern} J.~P.,  1999, \mn@doi [\apjl] {10.1086/312109}, \href {https://ui.adsabs.harvard.edu/abs/1999ApJ...519L..17S} {519, L17}

\bibitem[\protect\citeauthoryear{{Sault} \& {Wieringa}}{{Sault} \& {Wieringa}}{1994}]{Sault1994}
{Sault} R.~J.,  {Wieringa} M.~H.,  1994, \aaps, \href {https://ui.adsabs.harvard.edu/abs/1994A&AS..108..585S} {108, 585}

\bibitem[\protect\citeauthoryear{{Sault}, {Teuben}  \& {Wright}}{{Sault} et~al.}{1995}]{Sault1995}
{Sault} R.~J.,  {Teuben} P.~J.,   {Wright} M.~C.~H.,  1995, in {Shaw} R.~A.,  {Payne} H.~E.,   {Hayes} J.~J.~E.,  eds,  Astronomical Society of the Pacific Conference Series Vol. 77, Astronomical Data Analysis Software and Systems IV. p.~433 (\mn@eprint {arXiv} {astro-ph/0612759}), \mn@doi{10.48550/arXiv.astro-ph/0612759}

\bibitem[\protect\citeauthoryear{{Savchenko} et~al.,}{{Savchenko} et~al.}{2024}]{INTEGRAL2024}
{Savchenko} V.,  et~al., 2024, \mn@doi [A&A] {10.1051/0004-6361/202346336}, 684, L2

\bibitem[\protect\citeauthoryear{{Schulze} et~al.,}{{Schulze} et~al.}{2011}]{2011A&A...526A..23S}
{Schulze} S.,  et~al., 2011, \mn@doi [\aap] {10.1051/0004-6361/201015581}, \href {https://ui.adsabs.harvard.edu/abs/2011A&A...526A..23S} {526, A23}

\bibitem[\protect\citeauthoryear{{Shimwell} et~al.,}{{Shimwell} et~al.}{2019}]{Shimwell19}
{Shimwell} T.~W.,  et~al., 2019, \mn@doi [\aap] {10.1051/0004-6361/201833559}, \href {https://ui.adsabs.harvard.edu/abs/2019A&A...622A...1S} {622, A1}

\bibitem[\protect\citeauthoryear{{Shrestha} et~al.,}{{Shrestha} et~al.}{2023}]{Shrestha2023}
{Shrestha} M.,  et~al., 2023, \mn@doi [\apjl] {10.3847/2041-8213/acbd50}, \href {https://ui.adsabs.harvard.edu/abs/2023ApJ...946L..25S} {946, L25}

\bibitem[\protect\citeauthoryear{{Sironi}, {Spitkovsky}  \& {Arons}}{{Sironi} et~al.}{2013}]{2013ApJ...771...54S}
{Sironi} L.,  {Spitkovsky} A.,   {Arons} J.,  2013, \mn@doi [\apj] {10.1088/0004-637X/771/1/54}, \href {https://ui.adsabs.harvard.edu/abs/2013ApJ...771...54S} {771, 54}

\bibitem[\protect\citeauthoryear{{Starling}, {Wijers}, {Hughes}, {Tanvir}, {Vreeswijk}, {Rol}  \& {Salamanca}}{{Starling} et~al.}{2005}]{2005MNRAS.360..305S}
{Starling} R.~L.~C.,  {Wijers} R.~A.~M.~J.,  {Hughes} M.~A.,  {Tanvir} N.~R.,  {Vreeswijk} P.~M.,  {Rol} E.,   {Salamanca} I.,  2005, \mn@doi [\mnras] {10.1111/j.1365-2966.2005.09042.x}, \href {https://ui.adsabs.harvard.edu/abs/2005MNRAS.360..305S} {360, 305}

\bibitem[\protect\citeauthoryear{{Tanvir} et~al.,}{{Tanvir} et~al.}{2010}]{2010ApJ...725..625T}
{Tanvir} N.~R.,  et~al., 2010, \mn@doi [\apj] {10.1088/0004-637X/725/1/625}, \href {https://ui.adsabs.harvard.edu/abs/2010ApJ...725..625T} {725, 625}

\bibitem[\protect\citeauthoryear{{Tasse} et~al.,}{{Tasse} et~al.}{2021}]{Tasse21}
{Tasse} C.,  et~al., 2021, \mn@doi [\aap] {10.1051/0004-6361/202038804}, \href {https://ui.adsabs.harvard.edu/abs/2021A&A...648A...1T} {648, A1}

\bibitem[\protect\citeauthoryear{{Tiengo} et~al.,}{{Tiengo} et~al.}{2023}]{2023ApJ...946L..30T}
{Tiengo} A.,  et~al., 2023, \mn@doi [\apjl] {10.3847/2041-8213/acc1dc}, \href {https://ui.adsabs.harvard.edu/abs/2023ApJ...946L..30T} {946, L30}

\bibitem[\protect\citeauthoryear{{Vasilopoulos}, {Karavola}, {Stathopoulos}  \& {Petropoulou}}{{Vasilopoulos} et~al.}{2023}]{2023MNRAS.521.1590V}
{Vasilopoulos} G.,  {Karavola} D.,  {Stathopoulos} S.~I.,   {Petropoulou} M.,  2023, \mn@doi [\mnras] {10.1093/mnras/stad375}, \href {https://ui.adsabs.harvard.edu/abs/2023MNRAS.521.1590V} {521, 1590}

\bibitem[\protect\citeauthoryear{{Wijers} \& {Galama}}{{Wijers} \& {Galama}}{1999}]{wijers1999}
{Wijers} R.~A.~M.~J.,  {Galama} T.~J.,  1999, \mn@doi [\apj] {10.1086/307705}, \href {https://ui.adsabs.harvard.edu/abs/1999ApJ...523..177W} {523, 177}

\bibitem[\protect\citeauthoryear{{Williams} et~al.,}{{Williams} et~al.}{2016}]{Williams16}
{Williams} W.~L.,  et~al., 2016, \mn@doi [\mnras] {10.1093/mnras/stw1056}, \href {https://ui.adsabs.harvard.edu/abs/2016MNRAS.460.2385W} {460, 2385}

\bibitem[\protect\citeauthoryear{{Williams} et~al.,}{{Williams} et~al.}{2023}]{williams2023}
{Williams} M.~A.,  et~al., 2023, \mn@doi [\apjl] {10.3847/2041-8213/acbcd1}, \href {https://ui.adsabs.harvard.edu/abs/2023ApJ...946L..24W} {946, L24}

\bibitem[\protect\citeauthoryear{{Zwart} et~al.,}{{Zwart} et~al.}{2008}]{2008MNRAS.391.1545Z}
{Zwart} J.~T.~L.,  et~al., 2008, \mn@doi [\mnras] {10.1111/j.1365-2966.2008.13953.x}, \href {https://ui.adsabs.harvard.edu/abs/2008MNRAS.391.1545Z} {391, 1545}

\bibitem[\protect\citeauthoryear{{de Gasperin} et~al.,}{{de Gasperin} et~al.}{2019}]{deGasperin19}
{de Gasperin} F.,  et~al., 2019, \mn@doi [\aap] {10.1051/0004-6361/201833867}, \href {https://ui.adsabs.harvard.edu/abs/2019A&A...622A...5D} {622, A5}

\bibitem[\protect\citeauthoryear{{de Ugarte Postigo} et~al.,}{{de Ugarte Postigo} et~al.}{2022}]{2022GCN.32648....1D}
{de Ugarte Postigo} A.,  et~al., 2022, GRB Coordinates Network, \href {https://ui.adsabs.harvard.edu/abs/2022GCN.32648....1D} {32648}

\bibitem[\protect\citeauthoryear{{van Haarlem} et~al.,}{{van Haarlem} et~al.}{2013}]{2013A&A...556A...2V}
{van Haarlem} M.~P.,  et~al., 2013, \mn@doi [\aap] {10.1051/0004-6361/201220873}, \href {https://ui.adsabs.harvard.edu/abs/2013A&A...556A...2V} {556, A2}

\bibitem[\protect\citeauthoryear{{van Weeren} et~al.,}{{van Weeren} et~al.}{2016}]{vanWeeren16}
{van Weeren} R.~J.,  et~al., 2016, \mn@doi [\apjs] {10.3847/0067-0049/223/1/2}, \href {https://ui.adsabs.harvard.edu/abs/2016ApJS..223....2V} {223, 2}

\bibitem[\protect\citeauthoryear{{van der Horst}, {Rol}, {Wijers}, {Strom}, {Kaper}  \& {Kouveliotou}}{{van der Horst} et~al.}{2005}]{2005ApJ...634.1166V}
{van der Horst} A.~J.,  {Rol} E.,  {Wijers} R.~A.~M.~J.,  {Strom} R.,  {Kaper} L.,   {Kouveliotou} C.,  2005, \mn@doi [\apj] {10.1086/497021}, \href {https://ui.adsabs.harvard.edu/abs/2005ApJ...634.1166V} {634, 1166}

\bibitem[\protect\citeauthoryear{{van der Horst} et~al.,}{{van der Horst} et~al.}{2014}]{vanderhorst2014}
{van der Horst} A.~J.,  et~al., 2014, \mn@doi [\mnras] {10.1093/mnras/stu1664}, 444, 3151

\makeatother
\end{thebibliography}



\appendix

\bsp	
\label{lastpage}
\end{document}